\documentclass[11pt]{article}

\usepackage[latin1]{inputenc}
\usepackage{hyperref}
\usepackage{latexsym}
\usepackage{physics}
\usepackage{cancel}
\usepackage{braket}
\usepackage{times}
\usepackage{mathtools}
\usepackage{xcolor}
\usepackage{amsmath}
\usepackage{bbm}
\usepackage{yhmath}
\usepackage{amsthm}
\usepackage{leftidx}
\usepackage{amsfonts}
\usepackage{braket}
\usepackage{slashed}
\usepackage{subfigure}
\usepackage{soul}
\usepackage{cite}

\usepackage{dblfloatfix}    

\newenvironment{sciabstract}{%
\begin{quote} }
{\end{quote}}

\newcounter{lastnote}

\title{\huge{A Monte Carlo Approach to the Worldline Formalism in Curved Space}}

\author
{Olindo Corradini$\,$\footnote{E-mail: olindo.corradini@unimore.it} $\ $ and   Maurizio Muratori$\,$\footnote{E-mail: maurizio.muratori@unimore.it} \\
\\
\normalsize{Dipartimento di Scienze Fisiche, Informatiche e Matematiche,}\\ 
\normalsize{Universit\`a di Modena e Reggio Emilia, via Campi 213/A, I-41125 Modena, Italy}\\[2mm]
\normalsize{INFN, Sezione di Bologna, via Irnerio 46, I-40126 Bologna, Italy}
}

\date{\today}

\newcommand*{\defeq}{\mathrel{\vcenter{\baselineskip0.5ex \lineskiplimit0pt
                     \hbox{\scriptsize.}\hbox{\scriptsize.}}}%
                     =}
\begin{document}

\baselineskip15pt

\maketitle

\begin{sciabstract}
\begin{center}\bf{Abstract}\end{center}
We present a numerical method to evaluate worldline (WL) path integrals defined on a curved Euclidean space, sampled with Monte Carlo (MC) techniques. In particular, we adopt an algorithm known as \textit{YLOOPS} with a slight modification due to the introduction of a quadratic term which has the function of stabilizing and speeding up the convergence. Our method, as the perturbative counterparts, treats the non-trivial measure and deviation of the kinetic term from flat, as interaction terms. Moreover, the numerical discretization adopted in the present WLMC is realized with respect to the proper time of the associated bosonic point-particle, hence such procedure may be seen as an analogue of the time-slicing (TS) discretization already introduced to construct quantum path integrals in curved space. As a result, a TS counter-term is taken into account during the computation. The method is tested against existing analytic calculations of the heat kernel for a free bosonic point-particle in a $D$-dimensional maximally symmetric space.
\vspace*{\fill}
\end{sciabstract}

\section{Introduction}
\label{sec:intro}
The worldline formalism applied to Quantum Field Theory is a powerful tool to compute relevant physical quantities. The guidelines of such method were already introduced in the 1950 by Feynman, who proposed a quantum mechanical path integral model representing the dressed scalar propagator in scalar QED \cite{Feynman:1950ir}. In the 80's the method started to be systematically used as an alternative to standard second-quantized approaches to QFT, in particular for the computation of anomalies \cite{AlvarezGaume:1983at, AlvarezGaume:1983ig, Friedan:1983xr, Bastianelli:1991be, Bastianelli:1992ct}, effective actions and Feynman diagrams \cite{Bern:1990cu, Strassler:1992zr, Schubert:2001he, Edwards:2019eby}. This formalism was then extended to various quantum field theories in curved spacetime (see Ref.~\cite{Bastianelli:2006rx} for a review) and several applications have so far been considered such as, one-loop effective actions~\cite{Bastianelli:2002fv, Bastianelli:2002qw, Bastianelli:2005vk}, the one-loop graviton photon mixing in an electromagnetic field~\cite{Bastianelli:2004zp}, gravitational corrections to Euler-Heisenberg lagrangians~\cite{Bastianelli:2008cu}, worldline representations of quantum gravity~\cite{Bastianelli:2013tsa, Bonezzi:2018box}, supergravity~\cite{Bonezzi:2020jjq} and higher spin field theories~\cite{Bastianelli:2008nm, Bastianelli:2012bn}, and simplified methods for anomaly computations~\cite{Bastianelli:2001tb, Corradini:2018lov}, just to name a few.

The effort to move towards non-perturbative worldline calculations led to the development of numerical methods for the evaluation of quantum mechanical path integrals in flat space. The need for such a generalization was in the several interesting applications that rely on non-perturbative physics, such as the chiral symmetry breaking. The main issue was the numerical generation of the worldlines in an Euclidean space according to their relative probability distribution. For such a purpose, Gies and Langfeld \cite{Gies:2001zp} proposed a first numerical algorithm, adopting a Monte Carlo sampling of the coordinate space as an improvement of a previous work by Nieuwenhuis and Tjon \cite{Nieuwenhuis:1996mc}. Such methods are now known as \textit{Worldline Monte Carlo} (WLMC) and have been used for several application in QFT, such as the Casimir effect \cite{Gies:2003cv}, Schwinger pair production in inhomogeneous fields \cite{Gies:2005bz}, quantum effective actions~\cite{Gies:2005sb}, strongly-coupled, large-$N$ fermion models \cite{Dunne:2009zz}, {\color{black} and the nonperturbative propagator of scalar, QED-like, theories~\cite{Franchino-Vinas:2019udt}}. In these works, all the proposed algorithms are based on a Monte Carlo sampling which models the free Brownian motion. Thus, instead of using the full (kinetic + potential) action as a weight to select the points on the worldline, only the kinetic part is used. As explained in the recent work by Edwards et al.~\cite{Edwards:2019fjh}, the reason for such a choice is universality, which then allows the application of the method to a variety of different cases, regardless of the specific potential involved. Moreover, therein they propose an efficient algorithm, called \textit{YLOOPS}, to generate closed worldlines around a point,\footnote{Closed worldlines (loops) are due to quantum mechanical path integrals with periodic boundary conditions, often used to describe effective actions in the context of the worldline formalism.} which is an improved version of the alghoritms \textit{VLOOPS} and \textit{DLOOPS} originally designed in~\cite{Gies:2003cv} and~\cite{Gies:2005sb} respectively\footnote{\color{black}Actually, a parallelized implementation on GPUs of the \textit{DLOOPS} algorithm was considered by Aehlig et al. in~\cite{Aehlig:2011xg}.}. Below, due to its efficiency and easiness of application, we find it convenient to adopt the \textit{YLOOPS} algorithm for our calculation: more precisely, we present a slight modification of such algorithm, where a fictitious quadratic term is inserted in the kinetic term, in order to make the evaluation of the numerical path integral faster and more stable. 

The main goal of the present manuscript, is to extend WLMC techniques to curved spaces, i.e. to non-linear sigma models representing the propagation of a scalar particle in curved space. This non-perturbative numerical method would presumably help tackling a large class of problems in quantum field theory in curved space such as, for example, the Casimir effect and the Schwinger effect on a non trivial background, and the computation of effective actions and anomalies in curved space.  In order to test our construction we restrict ourselves to maximally symmetric spaces. In such a context indeed, recently it was shown that the heat kernel expansion for a scalar particle~\cite{Bastianelli:2017wsy, Guven:1987en} (later generalized to a spinor particle~\cite{Bastianelli:2018twb}) can be efficiently reproduced using an effective model with a flat kinetic term, where the effects of the curvatures are taken into account through a suitable effective potential. Therefore, as we show, such a model can be easily simulated with the conventional (flat space) WLMC methods, mentioned above. Moreover, this constitutes a perfect benchmark to verify our extension of the WLMC techniques to a genuine non-linear sigma model representing a particle in curved space.  

The paper is organized as follows. In Section \ref{sec:WLMC_FS} we review the Worldline Monte Carlo theory in flat space, providing an example of the calculation of the propagator associated to a 4-dimensional harmonic oscillator, to get familiar with the method. In Section \ref{sec:WLMC_CS} we build the WLMC setup in curved space and propose our strategy to compute numerical path integrals on non-trivial backgrounds. In Section \ref{sec:CS_fsHToS} we describe the theoretical basis of the case study which we will use to test our method, i.e. the diagonal part of the heat kernel of a free scalar point particle moving on a $D$-sphere and on a $D$-hyperboloid, i.e. maximally symmetric spaces. The worldline realization of such quantity is given in terms of a quantum mechanical path integrals with periodic boundary conditions (defining \textit{closed trajectories} or \textit{loops}). After that, we will report our numerical simulations, studying their convergence with respect to the discretization parameters and the curvature of the sphere. A possible extension to the case of \textit{open trajectories} is discussed in Section \ref{sec:openWL}. In Section \ref{sec:C} we sum up our conclusions, with possible outlook for future applications. Finally, in Appendix \ref{app:A} we report the details of the \textit{YLOOPS} algorithm generalized to the case of a non-vanishing quadratic term, whereas in Appendix \ref{app:B} we see the effect of inserting such mass term in the WL sampling discussed in Section \ref{sec:WLMC_FS}.
 
\section{Worldline Monte Carlo in flat space}
\label{sec:WLMC_FS}
In order to introduce our computation, let us first review the main ingredients of worldline Monte Carlo in flat space. Let us consider a $D$-dimensional flat space heat kernel computed through an Euclidean quantum mechanical path integral
\begin{align}
\label{eq1}
I(x,x';\beta)=\smashoperator{\int_{x(0)=x}^{x(\beta)=x'}} Dx\;e^{-S[x]},
\end{align}
with
\begin{align}
Dx=\smashoperator{\prod_{\tau=0,\ldots,\beta}}dx(\tau)
\end{align}
and
\begin{align}
\label{eq2}
S[x]=\int\displaylimits_0^{\beta}d\tau\left(\frac{1}{2}\delta_{\mu\nu}\dot x^{\mu}\dot x^{\nu}+V(x)\right).
\end{align} 
Expression \eqref{eq1} can be seen as the heat kernel of a unit mass non-relativistic point-particle $x(\tau)$ in $D$-dimensional Euclidean space, or conversely the (Schwinger integrand of the) propagator of a bosonic relativistic particle, where the affine parameter $\tau$ is the proper time. Its Hamiltonian reads
\begin{equation}
H[x,p]=\frac{1}{2}\delta^{\mu\nu}p_{\mu}p_{\nu}+V(x)
\end{equation}
The formal expression \eqref{eq1} involves the \textit{sum} over all the possible \textit{worldlines} joining $x$ and $x'$ in time $\beta$, weighted by the action \eqref{eq2}. To provide a numerical realization of the worldline path integral \eqref{eq1}, some comments are needed. First of all, it is not possible to span numerically the entire configuration space, hence a selection on the worldlines must be taken into account: let us denote $N_{WL}$ the number of worldlines which are considered. Secondly, each worldline has to be discretized: this is usually done with respect to the affine parameter, taking a number $N$ of points per worldline. As pointed out in \cite{Edwards:2019fjh}, here discretization is not realized on the points $x(\tau)$ of the manifold,\footnote{As done, for instance, in Lattice Field Theory.} but directly on the affine parameter $\tau$. To fulfill the above two requirements, what is usually done in WLMC approaches \cite{Gies:2001zp, Gies:2003cv, Gies:2005bz, Gies:2005sb, Edwards:2019fjh} is to compute \textit{worldline averages} like
\begin{align}
\label{eq3}
\left\langle I(x,x';\beta)\right\rangle=\frac{\displaystyle\int\displaylimits_{x(0)=x}^{x(\beta)=x'} \!\!\!\!\!\!Dx\;e^{-S_{KIN}[x]-S_{POT}[x]}}{\displaystyle\int\displaylimits_{x(0)=x}^{x(\beta)=x'} \!\!\!\!\!\!Dx\;e^{-S_{KIN}[x]}},
\end{align}
with $S_{KIN}[x]$ and $S_{POT}[x]$ being the kinetic and potential terms in \eqref{eq2} respectively. In this way, the exponential $e^{-S_{KIN}}$ characterizing each worldline can be used as a weight function to build the worldlines used for the calculation. To be more precise, worldlines are constructed point by point using Monte Carlo algorithms: each trajectory produced, then, satisfies a Monte Carlo-selection criterion which allows to consider all the trajectories obtained as a set of \textit{almost equally meaningful} trajectories to simulate the quantum propagation of the point-particle in spacetime. The aforementioned choice of the WL weight function is the most popular in literature, as it has an evident feature of \textit{universality}, in fact in such case the selection of the worldline is independent of the model considered. Once the configuration space is sampled with the points $\left\{ x^{(s)}_i \right\}_{i=1,\ldots,N}$, with $s=1,\ldots,N_{WL}$ denoting the worldline index, the potential can be in general approximated by\footnote{Such approximation works well with well-behaved potentials; when divergences appear, different approaches should be considered \cite{Edwards:2019fjh}.}
\begin{align}
S_{POT}^{(s)}(\beta)=\int\displaylimits_0^{\beta}d\tau\;V\left(x^{(s)}\right)\simeq\frac{\beta}{N}\sum_{i=1}^NV\left(x_i^{(s)}\right),\quad s=1,\ldots,N_{WL},
\end{align}
where all the worldlines start at $x$ and end at $x'$. At this point, expression \eqref{eq3} can be approximated by the arithmetic average
\begin{align}
\label{eq16}
\left\langle I(x,x';\beta)\right\rangle\simeq\frac{\displaystyle\sum_{s=1}^{N_{WL}}e^{-S_{POT}^{(s)}(\beta)}}{N_{WL}},
\end{align}
for which, clearly, a good estimate of the path integral is obtained when $N$ and $N_{WL}$ are large enough. 
Expression~\eqref{eq16} is the main prescription for our WLMC computations. Note that, unlike~\eqref{eq3}, the latter does not occur in the form of a weighted average. However, the weighing procedure takes place beforehand, upon the selection of the different worldlines present in the ensemble.   

As mentioned before, different algorithms have been developed in order to generate worldlines. However, historically, the most commonly used algorithm is the \textit{VLOOPS} developed by Gies \cite{Gies:2001zp}. Here we find it convenient to use the \textit{YLOOPS} routine developed by Edwards et al. \cite{Edwards:2019fjh}, with a slight modification due to the inclusion of a regularizing quadratic contribution in the kinetic term that will be taken into account for the computations in curved space. Such algorithm is developed to diagonalize the discretized version of the flat kinetic term in \eqref{eq2}, having vanishing Dirichlet boundary conditions $x_1=x_N=0$. The details are discussed in Appendix \ref{app:A}.

\subsection{A simple flat-space test}
In order to conclude this preliminary section let us review a simple example where the WLMC setup can be applied, i.e. the propagator of a bosonic harmonic oscillator with null endpoints in a $D$-dimensional Euclidean space. Such quantity can be written as
\begin{equation}
\label{eq4}
\tilde G(\beta)\defeq\smashoperator{\int_{x(0)=0}^{x(\beta)=0}}Dx\;e^{-\frac{1}{2}\int_0^{\beta}d\tau\left( \dot x^2+x^2 \right)},
\end{equation}
where we have assumed $\omega=1=m$. The analytical result is well known and reads
\begin{equation}
\label{eq5}
\tilde G(\beta)=\left[ \frac{1}{2\pi\sinh(\beta)} \right]^{\frac{D}{2}}.
\end{equation}
On the other hand, the propagator can be written as
\begin{equation}
\label{eq6}
\tilde G(\beta)=\oint_0 Dx\;e^{-\frac{1}{2}\int_0^{\beta}d\tau\; \dot x^2}\;\frac{\displaystyle\oint_0 Dx\;e^{-\frac{1}{2}\int_0^{\beta}d\tau\left( \dot x^2+x^2 \right)}}{\displaystyle\oint_0 Dx\;e^{-\frac{1}{2}\int_0^{\beta}d\tau\; \dot x^2}}
\end{equation}
where we have multiplied and divided by the free path integral, which provides the Feynman factor $\left(\frac{1}{2\pi\beta}\right)^{\frac{D}{2}}$---above we denote $\oint_0:=\int_{x(0)=0}^{x(\beta)=0}$. Expression~\eqref{eq6} thus reads
\begin{equation}
\label{eq7}
\tilde G(\beta)=\left(\frac{1}{2\pi\beta}\right)^{\frac{D}{2}}\left\langle e^{-\frac{1}{2}\int_0^{\beta}d\tau\; x^2} \right\rangle
\end{equation}
and, combining \eqref{eq7} with \eqref{eq5}, we obtain
\begin{equation}
\label{eq8}
G(\beta)\defeq\left\langle e^{-\frac{1}{2}\int_0^{\beta}d\tau\; x^2} \right\rangle=\left[ \frac{\beta}{\sinh(\beta)} \right]^{\frac{D}{2}},
\end{equation}
which is accessible to WLMC calculation.
\begin{figure}[h!]
\includegraphics[width=\textwidth]{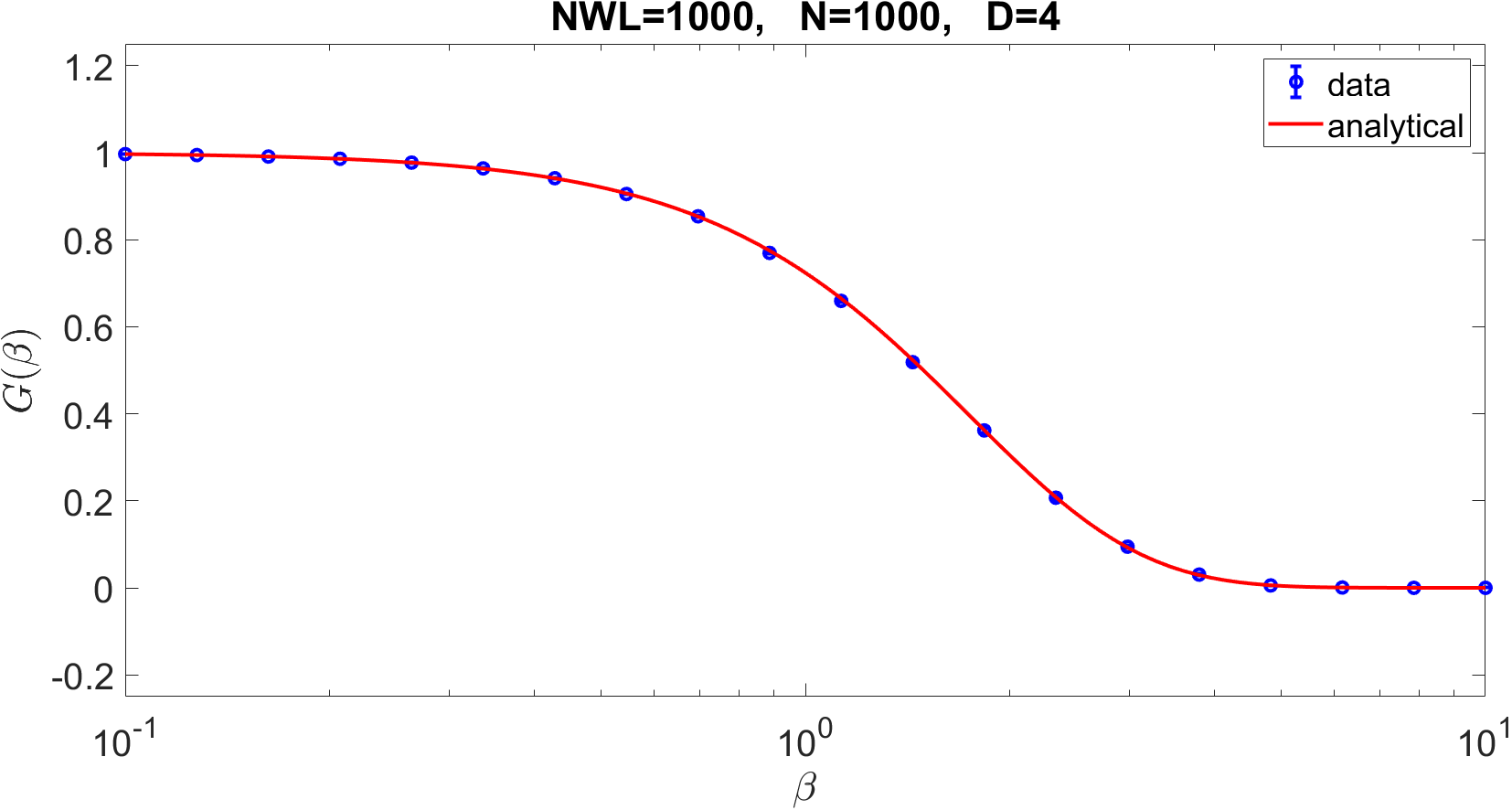}
\caption{Comparison between a WLMC simulation with $N=1000$, $N_{WL}=1000$ and the analytical result for the propagator of the $4$-dimensional bosonic harmonic oscillator with endpoints fixed at zero.}
\label{fig1}
\end{figure}
Indeed, in Fig.~\ref{fig1} we report a WLMC calculation of $G(\beta)$ in four Euclidean dimensions, with $N=1000$ points per loop and $N_{WL}=1000$ worldlines. It shows a satisfactory agreement with the analytical result in \eqref{eq8}.  In order to remove correlation between the data, this calculation---as well as those presented henceforth---has been performed using a different set of worldlines for each $\beta$-value.

{\color{black}Here and throughout the whole manuscript, the uncertainty on the numerical data is computed considering the standard error of the mean

\begin{equation}
    SEM=\sqrt{\sum_{s=1}^{N_{WL}}\frac{\left[(\ldots)_s-\left\langle (\ldots) \right\rangle  \right]^2}{N_{WL}(N_{WL}-1)}}
\end{equation}
where the dots represent a quantity computed for each worldline~\cite{Edwards:2019fjh}. 
}

As we have seen, WLMC in flat space is relatively easy to implement and, even for not very large values of $N$ and $N_{WL}$, it faithfully reproduces the theoretical curve. We will show that the flat space WLMC setup can be fruitfully exploited also for numerical applications in curved space problems: the price to pay is the introduction of further potential-like terms that incorporates curvature effects.  

\section{Worldline Monte Carlo in curved space}
\label{sec:WLMC_CS}
Similarly to what we did in the previous section for the flat space case, let us consider the classical  Hamiltonian of a $D$-dimensional bosonic scalar point-particle in Euclidean curved space
\begin{equation}
\mathcal H(x,p)=\frac{1}{2}g^{\mu\nu}(x)p_{\mu}p_{\nu}+V(x).
\end{equation}
At the quantum level the Einstein invariant Hamiltonian operator reads (see~\cite{Bastianelli:2006rx} for a complete treatment of particle path integrals in curved space)
\begin{equation}
\label{eq9}
\hat{\mathcal H}(\hat x, \hat p)=\frac{1}{2}g^{-\frac{1}{4}}(\hat x)\hat p_{\mu}g^{\frac{1}{2}}(\hat x)g^{\mu\nu}(\hat x)\hat p_{\nu}g^{-\frac{1}{4}}(\hat x) + V(\hat x),
\end{equation}
with $g(x)=\det g_{\mu\nu}(x)$, and the heat kernel associated to \eqref{eq9} can be expressed as the following particle path integral
\begin{equation}
I(x,x';\beta)=\smashoperator{\int_{x(0)=x}^{x(\beta)=x'}}\mathcal Dx\;e^{-S[x]},
\end{equation}
where the Einstein-invariant formal measure reads
\begin{equation}
\label{eq10}
\mathcal Dx=\smashoperator{\prod_{\tau=0,\ldots,\beta}}\sqrt{g\left(x(\tau)\right)}dx(\tau)
\end{equation}
and the resulting particle action 
\begin{equation}
\label{eq11}
S[x]=\int\displaylimits_0^{\beta}d\tau\left( \frac{1}{2}g_{\mu\nu}(x)\dot x^{\mu} \dot x^{\nu}+V_{CT}(x)+V(x) \right)\,,
\end{equation}
is a one-dimensional non-linear sigma model.

With respect to the flat space case, there are three main differences: the root factor in \eqref{eq10}, the spacetime dependent metric tensor $g_{\mu\nu}(x)$ and the counter-term $V_{CT}(x)$ appearing in \eqref{eq11}. A few comments are in order. In perturbative computations, one needs a regularization scheme to treat divergent Feynman diagrams and  $V_{CT}(x)$ is the suitable, finite, counterterm needed, which depends on the adopted regularization scheme. On the other hand, in non-perturbative computations, no divergences are expected to occur since these Feynman path integrals represent quantum mechanical transition amplitudes which are thus finite. {\color{black} Strictly speaking this a formal statement which boils down to a quite non-obvious definition of the path integral measure, so there is {\it a priori} no guarantee that divergences do not occur in intermediate stages of the computation. On the other hand, in the present approach, which makes use of Monte Carlo samplings to define the ensemble of paths which are included, no divergences are seen to show up.} 

{\color{black} In general, in curved space particle models}, there are ordering ambiguities as the kinetic hamiltonian mixes coordinates and momenta. Yet, a well known scheme that allows to obtain Feynman path integrals from the associated quantum mechanical transition amplitudes exists: it is the Time Slicing approach, which requires to Weyl-order the Einstein-inviariant hamiltonian, in order to employ the so-called ``mid-point rule'' in the costruction of the path integral. Such reordering produces an order-$\hbar$ potential, $V_{CT}(x)$, which is the same potential needed at the perturbative level, since the Time Slicing also encodes a regularization procedure. For historical reasons we will refer to such potential as the ``counterterm potential''.  Finally, the root factor in \eqref{eq10} renders the formal measure Einstein invariant: for perturbative calculations it is often customary to exponentiate the $\sqrt{g(x)}$ in terms of a particle path integrals over Lee-Yang ghost fields. However, for our non-perturbative, numeric computation we find it more convenient to keep it as an external factor, evaluated at each point of the various worldlines involved in the sums.

Now, the key idea in order to perform a numerical WLMC average in curved space (like the one in~\eqref{eq3}) is to extract the flat space kinetic term from the curved one
\begin{equation}
\label{eq13}
g_{\mu\nu}(x)\dot x^{\mu}\dot x^{\nu}=\delta_{\mu\nu}\dot x^{\mu}\dot x^{\nu}+\left(g_{\mu\nu}(x)-\delta_{\mu\nu}\right)\dot x^{\mu}\dot x^{\nu}
\end{equation}
and use it to sample the worldlines, and treat the remaining part as a potential contribution, so that it is possible to bring the curved space problem back to a flat space one---provided one takes into account the suitable counterterm and measure factors. Namely, after a flat space Monte Carlo sampling of the worldlines, the path integral average will thus be given by\begin{equation}
\label{eq17}
\left\langle I(x,x';\beta) \right\rangle\simeq\frac{\displaystyle\sum_{s=1}^{N_{WL}}\sqrt{g^{(s)}}\;e^{-S^{(s)}_{POT}(\beta)}}{N_{WL}},
\end{equation}
where
\begin{align}
\label{eq12}
\sqrt{g^{(s)}}&\defeq\prod\displaylimits_{n=2}^{N}\sqrt{g\left(\bar x^{(s)}_{n-\frac{1}{2}}\right)}\,,
\nonumber
\\
S^{(s)}_{POT}(\beta)&\defeq\frac{\beta}{N-1}\sum\displaylimits_{n=2}^N\left[ V_{KIN}(\bar x^{(s)}_{n-\frac{1}{2}})+V_{CT}(\bar x^{(s)}_{n-\frac{1}{2}})+V(\bar x^{(s)}_{n-\frac{1}{2}}) \right]~,
\end{align}
and 
\begin{align}
	x^{(s)}_{n-\frac{1}{2}}\defeq\frac{x^{(s)}_{n}+x^{(s)}_{n-1}}{2}
\end{align}
is the middle point between adjacent points of the worldline $s$.

In particular, the first potential term appearing in \eqref{eq12} is the one arising from \eqref{eq13} and reads
\begin{align}
\label{eq15}
V_{KIN}(\bar x^{(s)}_{n-\frac{1}{2}})\defeq\frac{\displaystyle (N-1)^2}{\displaystyle\beta^2}\left( g_{\mu\nu}(\bar x^{(s)}_{n-\frac{1}{2}})-\delta_{\mu\nu} \right)\left( x^{(s)\;\mu}_{n}-x^{(s)\;\mu}_{n-1} \right)\left( x^{(s)\;\nu}_{n}-x^{(s)\;\nu}_{n-1} \right).
\end{align}
It is easy to verify that, as discussed in the Appendix~\ref{app:A}, definitions \eqref{eq12} and \eqref{eq15} respect the backward convention on the first derivative. Focusing on the counter-term potential $V_{CT}$, it ought to be observed that the proper time discretization of the path integral which we performed is equivalent to the Time Slicing (TS) procedure adopted to regularize particle path integrals in curved space from first principles \cite{Bastianelli:2006rx}. Thus, the discretized counter-term is~\footnote{Our conventions for the curvature tensors are $[\nabla_\mu,\nabla_\nu]V^\lambda =R_{\mu\nu}{}^\lambda{}_{\sigma}V^\sigma$,  $R_{\mu\nu}=R_{\mu}{}^{\rho}{}_{\nu\rho}$.}
\begin{equation}
\label{eq21}
V_{TS}(\bar x^{(s)}_{n-\frac{1}{2}})=-\frac{1}{8}\left[ R\left(\bar x^{(s)}_{n-\frac{1}{2}}\right)+g^{\mu\nu}\Gamma^{\rho}{}_{\mu\sigma}\Gamma^{\sigma}{}_{\nu\rho}\left(\bar x^{(s)}_{n-\frac{1}{2}}\right) \right].
\end{equation}
Now we have all the ingredients to test our setup. 

In the following section we consider a case study which comes particularly handy for our purpose: it is the study of the free heat kernel of a scalar point particle constrained on a $D$-sphere. Recently, in Ref.~\cite{Bastianelli:2017wsy} (see also~\cite{Guven:1987en}), it was shown that, using Riemann normal coordinates (RNC), it is possible to map the problem of computing the heat kernel on spheres, to the evaluation of a flat space heat kernel with a suitable potential which takes into account the curvature effects. Moreover, as found in~\cite{Bastianelli:2001tb}, for a maximally symmetric geometry, the metric tensor, in RNC, can be neatly written in closed form as the flat space metric plus a curvature-dependent contribution. Hence, our numerical problem can be studied both from a purely flat perspective, by mapping it into the effective flat space model of~\cite{Bastianelli:2017wsy}---thus using the conventional flat space Worldline Monte Carlo reviewed in Section~\ref{sec:WLMC_FS}---, and using the curved space WLMC setup discussed in Section~\ref{sec:WLMC_CS}. In other words, we use the effective flat space model as a benchmark test for our curved construction.

\section{Case study: free scalar heat kernel on maximally symmetric spaces}
\label{sec:CS_fsHToS}
Let us briefly review the model \cite{Bastianelli:2017wsy} which we are going to use to implement our numerical WLMC method in curved space. The Hamiltonian operator 
\begin{equation}
\hat{\mathcal H}_0(\hat x, \hat p)=\frac{1}{2}g^{-\frac{1}{4}}(\hat x)\hat p_{\mu}g^{\frac{1}{2}}(\hat x)g^{\mu\nu}(\hat x)\hat p_{\nu}g^{-\frac{1}{4}}(\hat x)
\end{equation}
of a free scalar point-particle in curved space can be used to build its heat kernel
\begin{equation}
\hat{\mathcal K} (\beta)=e^{-\beta\hat{\mathcal H}_0}
\end{equation}
satisfying the operatorial equation
\begin{equation}
-\frac{\partial \hat{\mathcal K} (\beta)}{\partial\beta}=\hat{\mathcal H}_0\hat{\mathcal K} (\beta)~.
\end{equation}
The matrix elements of the kernel operator
\begin{equation}
\mathcal K(x,x';\beta)=\bra{x}\hat{\mathcal K}(\beta)\ket{x'}
\end{equation}
satisfy the heat equation
\begin{equation}
\label{eq18}
\frac{\partial \mathcal K(x,x';\beta)}{\partial\beta}=\frac{1}{2}\nabla_x{}^2\mathcal K(x,x';\beta)~.
\end{equation}
Defining
\begin{equation}
\bar{\mathcal K}(x,x';\beta)=g^{\frac{1}{4}}(x)\mathcal K(x,x';\beta)g^{\frac{1}{4}}(x')~,
\end{equation}
and using RNC, the curved space heat equation~\eqref{eq18} can be turned into the flat space heat equation
\begin{equation}
\label{eq19}
\frac{\partial \bar{\mathcal K}(x,x';\beta)}{\partial\beta}=\left(-\frac{1}{2}\delta^{\mu\nu}\partial_{\mu}\partial_{\nu}+V_{eff}(x)\right)\bar{\mathcal K}(x,x';\beta)
\end{equation}
if the curved space is a space with maximally symmetry (a $D$-sphere for definiteness), for which
\begin{align}
R_{\mu\nu\rho\sigma}=&\;M^2\left(g_{\mu\rho}g_{\nu\sigma}-g_{\mu\sigma}g_{\nu\rho}\right)~.
\end{align}
In such a case we have
\begin{align}
g_{\mu\nu}(x)=&\;\delta_{\mu\nu}+f(x)P_{\mu\nu}(x),\quad x=\sqrt{\delta_{\mu\nu}x^{\mu}x^{\nu}}\nonumber \\
f(x)=&\;\frac{1-2M^2x^2-\cos(2Mx)}{2M^2x^2},\quad P_{\mu\nu}(x)=\delta_{\mu\nu}-\frac{x_{\mu}x_{\nu}}{x^2}\nonumber\\
g(x)=&\;\left( 1+f(x) \right)^{D-1}~.
\label{eq34}
\end{align}
Moreover, the effective potential appearing in \eqref{eq19} can be expressed in closed form as 
\begin{equation}
\label{eq20}
V_{eff}(x)=\frac{D(1-D)}{12}M^2+\frac{(D-1)(D-3)}{48}\frac{5M^2x^2-3+\left( M^2x^2+3 \right)\cos(2Mx)}{x^2\sin^2(Mx)}~,
\end{equation}
and can be used in the associated path integral representation of the problem
\begin{equation}
\bar{\mathcal K}(x,x';\beta)=\smashoperator{\int_{x(0)=x}^{x(\beta)=x'}}Dx\;e^{-S[x]},\quad S[x]=\int_0^{\beta}d\tau\left( \frac{1}{2}\delta_{\mu\nu}\dot x^{\mu}\dot x^{\nu}+V_{eff}(x) \right).
\end{equation}
Hence, the heat kernel for a particle on a maximally symmetric space can be obtained through a linear sigma model with an effective potential that encodes the curvature effects. We use such effective representation of the heat kernel to test our numerical method.  

\begin{table}[ht!]
\centering
{\renewcommand\arraystretch{3}
\begin{tabular}{c|c|c}
 & Effective Potential Method (EPM) & Non-linear Sigma Model Method (NSMM)\\
\hline
WL algorithm & original \textit{YLOOPS} & generalized (massive) \textit{YLOOPS}\\
quantity & $\left\langle I(0,0;\beta) \right\rangle\simeq\frac{\displaystyle\sum_{s=1}^{N_{WL}}e^{-S_{POT}^{(s)}(\beta)}}{N_{WL}}$ & $\left\langle I(0,0;\beta) \right\rangle\simeq\frac{\displaystyle\sum_{s=1}^{N_{WL}}\sqrt{g^{(s)}}\;e^{-S^{(s)}_{POT}(\beta)}}{N_{WL}}$ \\
WL action & $S_{POT}^{(s)}(\beta)=\frac{\beta}{N}\displaystyle\sum_{n=1}^NV_{eff}(x_n^{(s)})$ & $S^{(s)}_{POT}(\beta)=\frac{\beta}{N-1}\displaystyle\sum_{n=2}^N\left[ V_{KIN}(\bar x^{(s)}_{n-\frac{1}{2}})+V_{TS}(\bar x^{(s)}_{n-\frac{1}{2}})\right]$\\
potential & eq. \eqref{eq20} & eqs. \eqref{eq15} and \eqref{eq21}
\end{tabular}}
\caption{Details of the two types of computation performed for the same heat kernel expansion, with a linear sigma model and with a non-linear sigma model respectively.}
\label{tab1}
\end{table}

Specifically, we summarize in Table~\ref{tab1} the details of the two numerical calculations that we have performed.  The first one, which we refer to as ``Effective Potential Method'' (EPM) makes use of the potential given in Eq.~\eqref{eq20} to effectively reproduce the heat kernel of a scalar particle on a maximally symmetric space through a linear sigma model (flat metric). In the second calculation, we compute the same heat kernel making use of a ``genuine'' non-linear sigma model action, which amounts to consider the two potentials \eqref{eq15} and \eqref{eq21} within the WLMC technique---we refer to this case as the ``Non-linear Sigma Model Method'' (NSMM). Note that, for the EPM computation we have adopted the original \textit{YLOOPS} algorithm, developed in \cite{Edwards:2019fjh}, whereas for the NSMM we found it convenient to use a generalized version of \textit{YLOOPS}, where a tiny quadratic term is included in the kinetic part of the action. Namely,
\begin{equation}
    S_{\text{KIN}}[x]=\frac{1}{2}\delta_{\mu\nu}\dot x^{\mu}\dot x^{\nu}\to \tilde S_{\text{KIN}}[x]=\frac{1}{2}\delta_{\mu\nu}\dot x^{\mu}\dot x^{\nu}+\frac{1}{2}\alpha x^2,\; 0<\alpha\ll1.
\end{equation}
Let us stress that such mass term is used solely at the stage of the construction of the worldline ensemble, to improve the convergence, and does not contribute to the potential term, which is left untouched. The details of the generalized \textit{YLOOPS} algorithm are discussed in Appendix~\ref{app:A}, whereas its numerical outcomes are shown in Appendix~\ref{app:B}, {\color{black} where an analysis of the effect of the inclusion of the fictitious mass term is provided. Our findings show that there is range of values of the mass parameter $\alpha$ which minimize the discrepancy between the two methods, at large values of $\beta$.}

\begin{figure}[ht!]
     \begin{center}
	   \subfigure[]{%
            \label{fig:100_100_prop}
            \includegraphics[width=0.5\textwidth]{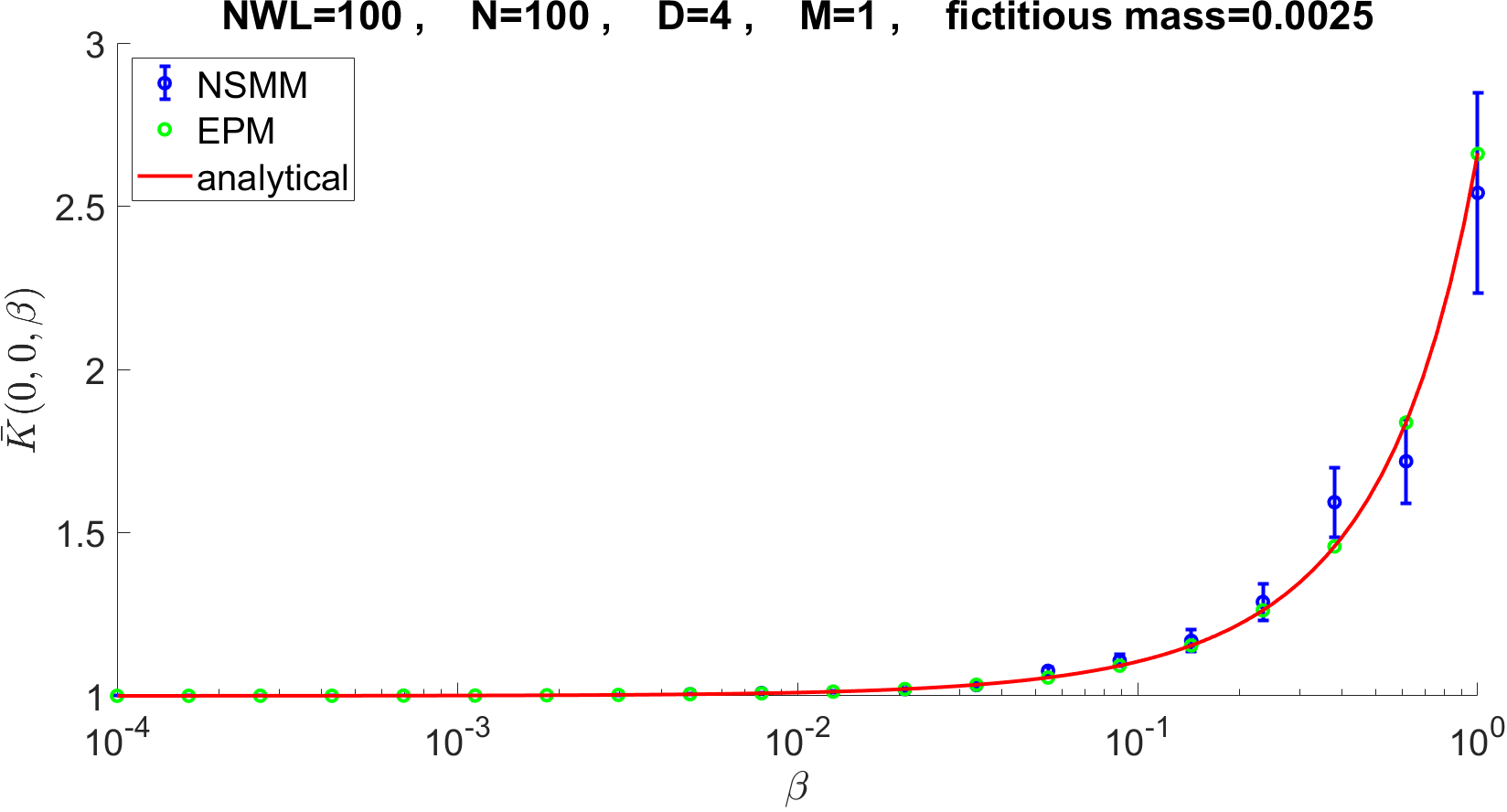}
        }%
        \subfigure[]{%
            \label{fig:3000_3000_prop}
            \includegraphics[width=0.5\textwidth]{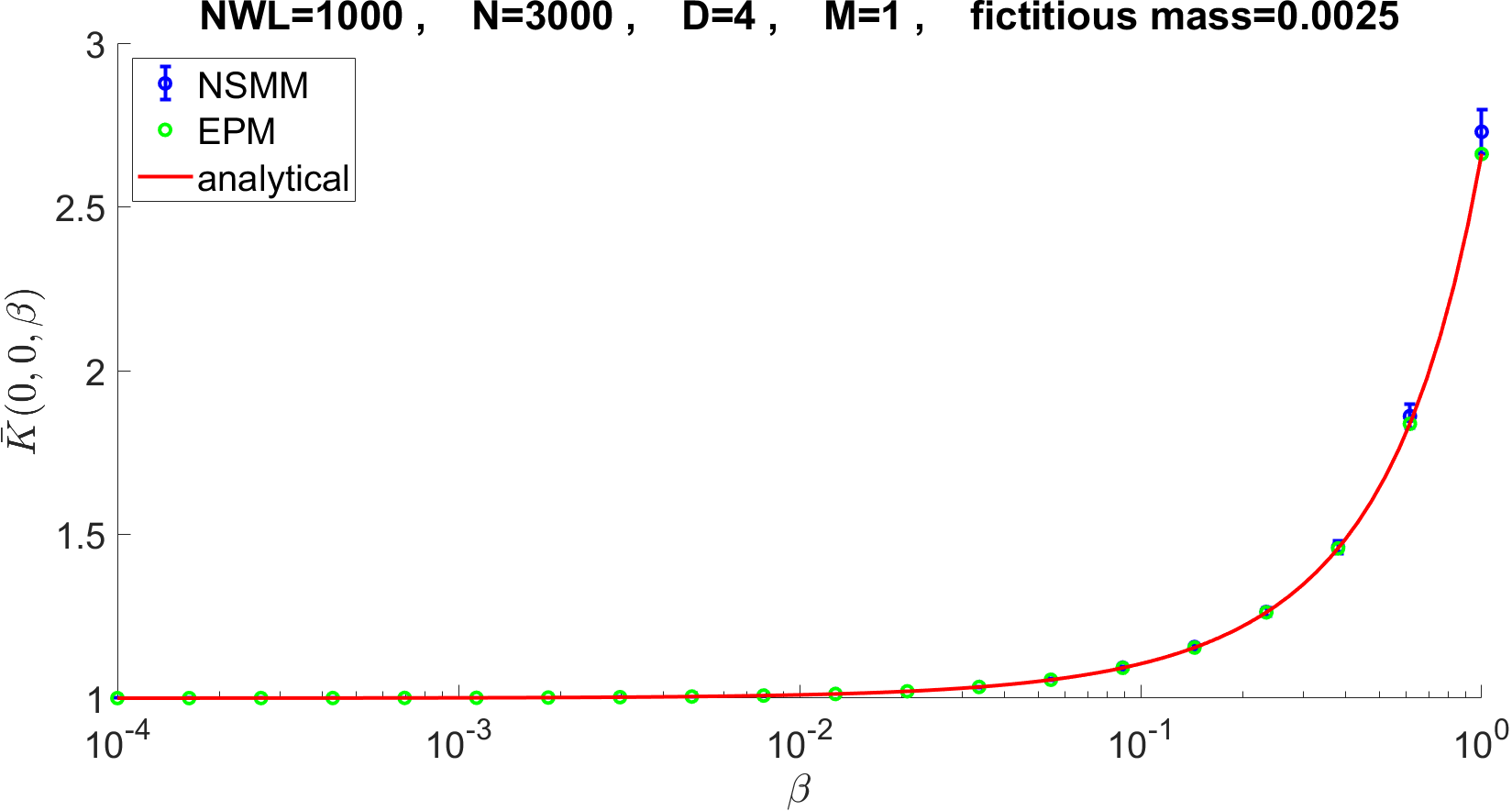}
        }%
%
    \end{center}
    \caption{%
        Calculations of the heat kernel of a free scalar particle in a $D$=4 sphere with endpoints $x=x'=0$ for various parameters $(N_{WL},N)$ both using the effective potential method (green data) with known potential and using the non-linear sigma model method (blue data). The red curve is the analytical and perturbative computation for small $\beta$ performed in \cite{Bastianelli:2017wsy}. 
     }%
   \label{fig:prop_relerr}
\end{figure}

In Figure \ref{fig:prop_relerr}, we report the results of our calculations for a $D=4$ sphere with $M=1$. Figures \ref{fig:100_100_prop} and \ref{fig:3000_3000_prop} show a comparison between the data of the Non-linear Sigma Model Method (NSMM, blue) and the Effective Potential Method (EPM, green), together with the perturbative, analytical computation performed in \cite{Bastianelli:2017wsy}, for couples of values $(N_{WL},N)=(100,100)$ and $(1000,3000)$ respectively. As we can see, the EPM is considerably more precise than the NSMM, presumably because of the form of the associated potential: in the first case it is expressed in \textit{regular} form \eqref{eq20}, whilst in the second case by means of numerical derivatives \eqref{eq15}, and this contributes to reduce precision. Thus, we take the green points as a benchmark to test the blue ones. We notice that the agreement between the two calculations gets better as the discretized approximation improves, i.e. when $N_{WL}$ and $N$ become large. Indeed, the absolute value of the maximum relative error passes from $\sim35\%$ with $(N_{WL},N)=(100,100)$ (fig. \ref{fig:100_100_prop}) to $\sim4\%$  with $(N_{WL},N)=(3000,3000)$ (fig. \ref{fig:3000_3000_prop}). In fact, it was noticed that WLMC calculations generally lose accuracy at large time values due to the numerical phenomenon of \textit{undersampling} \cite{Edwards:2019fjh} of worldlines. {\color{black}This issue is also studied in \cite{Gies:2005sb} for the specific case of the harmonic oscillator. It  would seem that our results confirm this feature, even though a source of systematic deviation from the reference curve has to be ascribed to a minor efficiency in representing the derivative interactions.} In the attempt to increase the efficiency of the code, one may think of using a higher order discrete derivative to insert into \eqref{eq15}, like the two-point method
\begin{equation}
\frac{f(x+h)-f(x-h)}{2h}=\frac{df}{dh}(x)+o(h^2)
\end{equation}
or, say, the five-point method
\begin{equation}
\frac{-f(x+2h)+8f(x+h)-8f(x-h)+f(x-2h)}{12h}=\frac{df}{dh}(x)+o(h^4).
\end{equation}
However, the specific form of the counter-term \eqref{eq21} is compatible only with a first order backward derivative, as it was derived \cite{Bastianelli:2006rx} in this way, and numerical simulations confirmed that.
\begin{figure}[h!]
   \centering
            \includegraphics[width=0.7\textwidth]{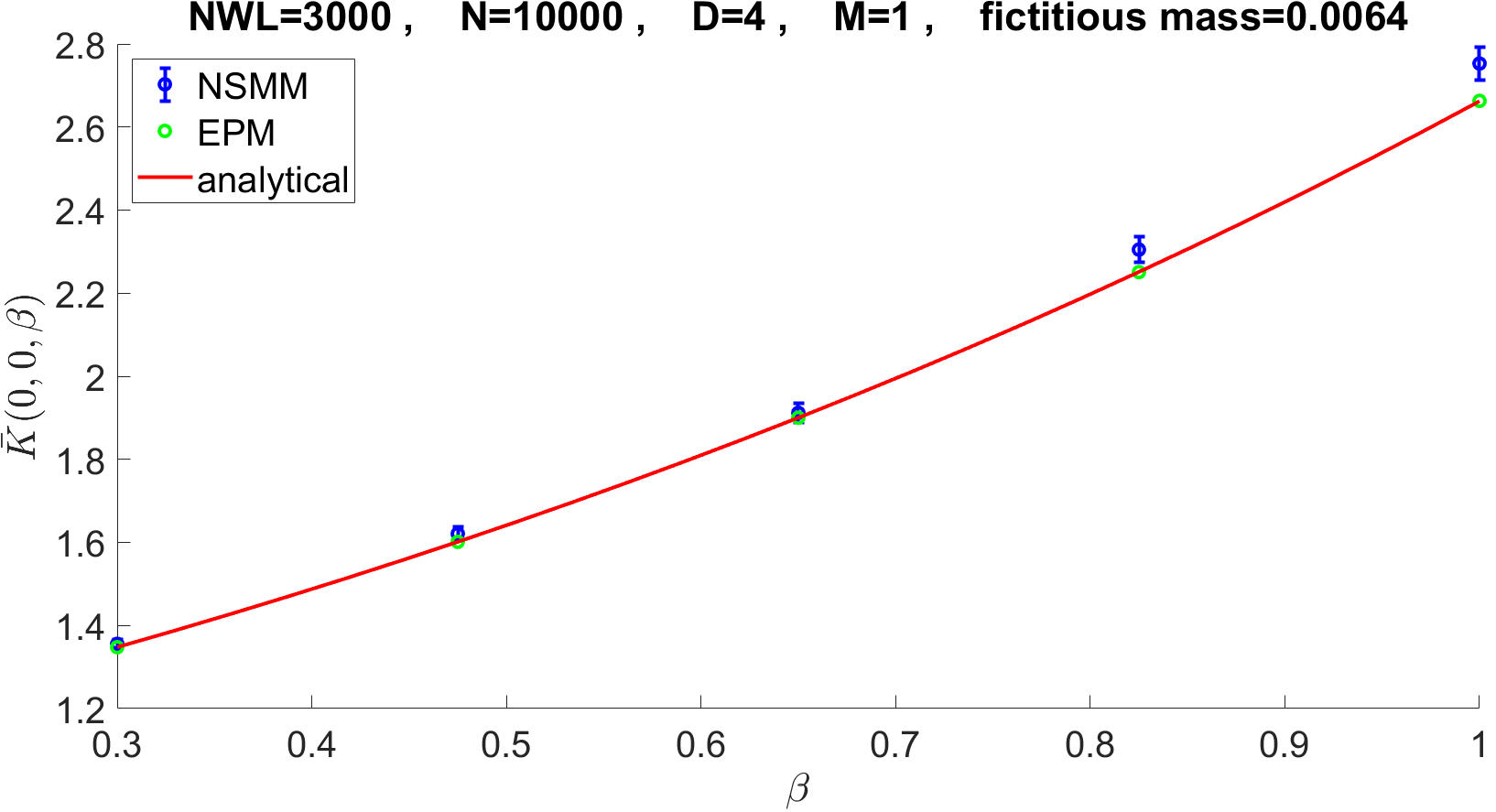}
    \caption{%
        High precision calculation of the heat kernel for the free scalar particle on the sphere with $(N_{WL},N)=(3000,10000)$.}
     \label{fig:prop_relerr_hp}%
\end{figure}

\begin{figure}[h!]
   \centering
            \includegraphics[width=0.7\textwidth]{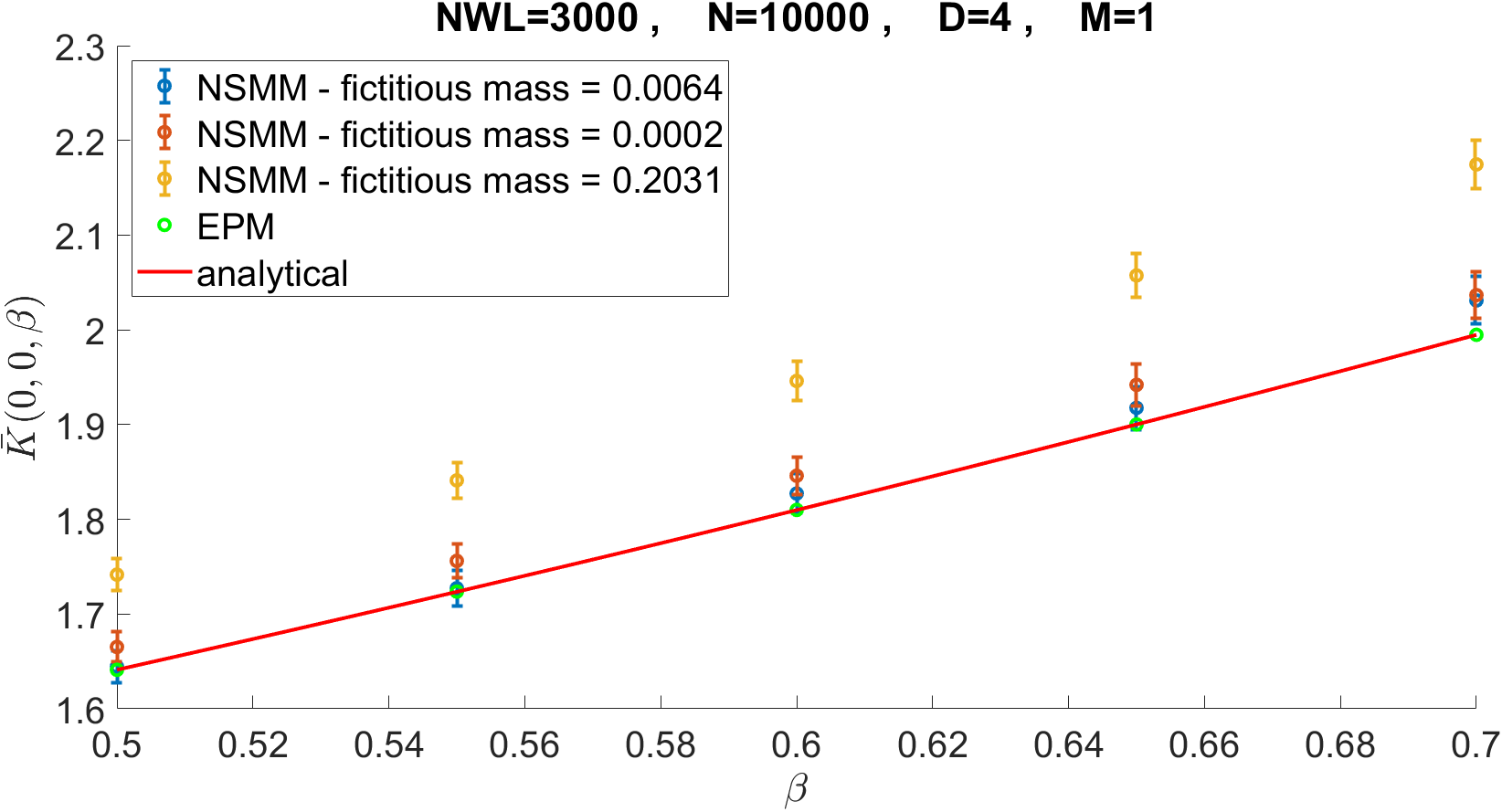}
    \caption{%
       Calculations of the heat kernel for the free scalar particle on the sphere with $(N_{WL},N)=(3000,10000)$ and or different values of the fictitious mass.}
     \label{fig:hp1}%
\end{figure}

In Figure~\ref{fig:prop_relerr_hp} we report a higher precision calculation with respect to those of Figures \ref{fig:prop_relerr}, in particular with parameters $(N_{WL},N)=(3000,10000)$ and for the interval $(0.3,1)$ of $\beta$-values, i.e. the tails of the Figures \ref{fig:prop_relerr}. Here we confirm a consistent suppression of Monte Carlo fluctuations, due to the consistent increase of the parameters $N_{WL}$ and $N$. {\color{black}To have a deeper insight of the effect of the fictitious mass $\alpha$, in Figure~\ref{fig:hp1} we report a higher precision calculation performed with  a set of three different values of $\alpha$, \textit{i.e.} $\alpha_{min}=0.0064$, $\alpha_{-}=0.0002$ and $\alpha_{+}=0.2031$. The choice of $\alpha_{min}$ is actually explained in Appendix \ref{app:B}: it represents the optimized value of $\alpha$ which minimizes the average discrepancy between the NSMM-data and the EPM-data. $\alpha_{-}$ and $\alpha_{+}$ lie respectively on the left and on the right of $\alpha_{min}$ on a logarithmic scale.}

\begin{figure}[ht!]
   \centering
        \subfigure[]{%
            \label{fig:M_1_prop}
            \includegraphics[width=0.5\textwidth]{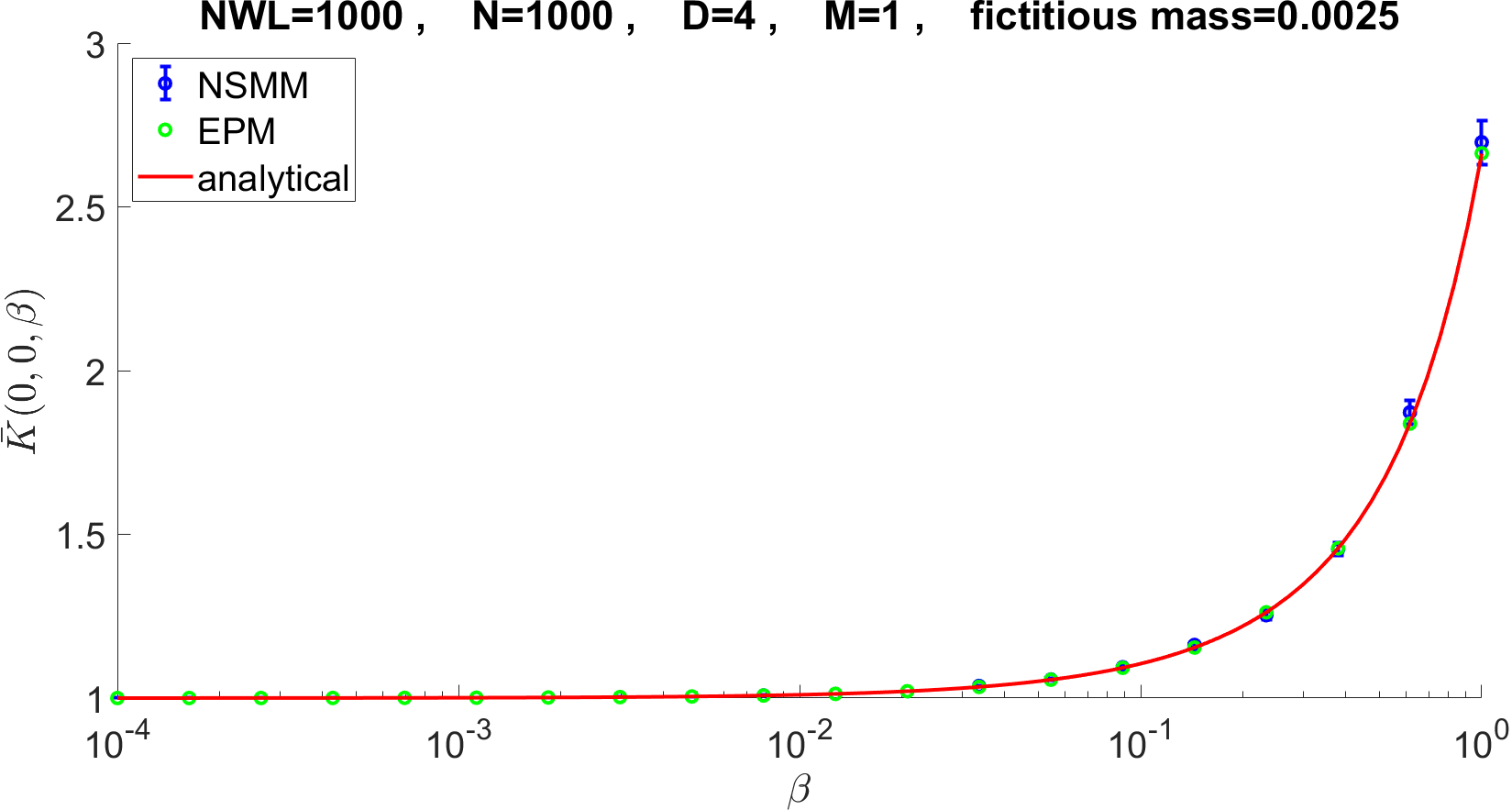}
        }%
        \subfigure[]{%
           \label{fig:M_1_relerr}
           \includegraphics[width=0.5\textwidth]{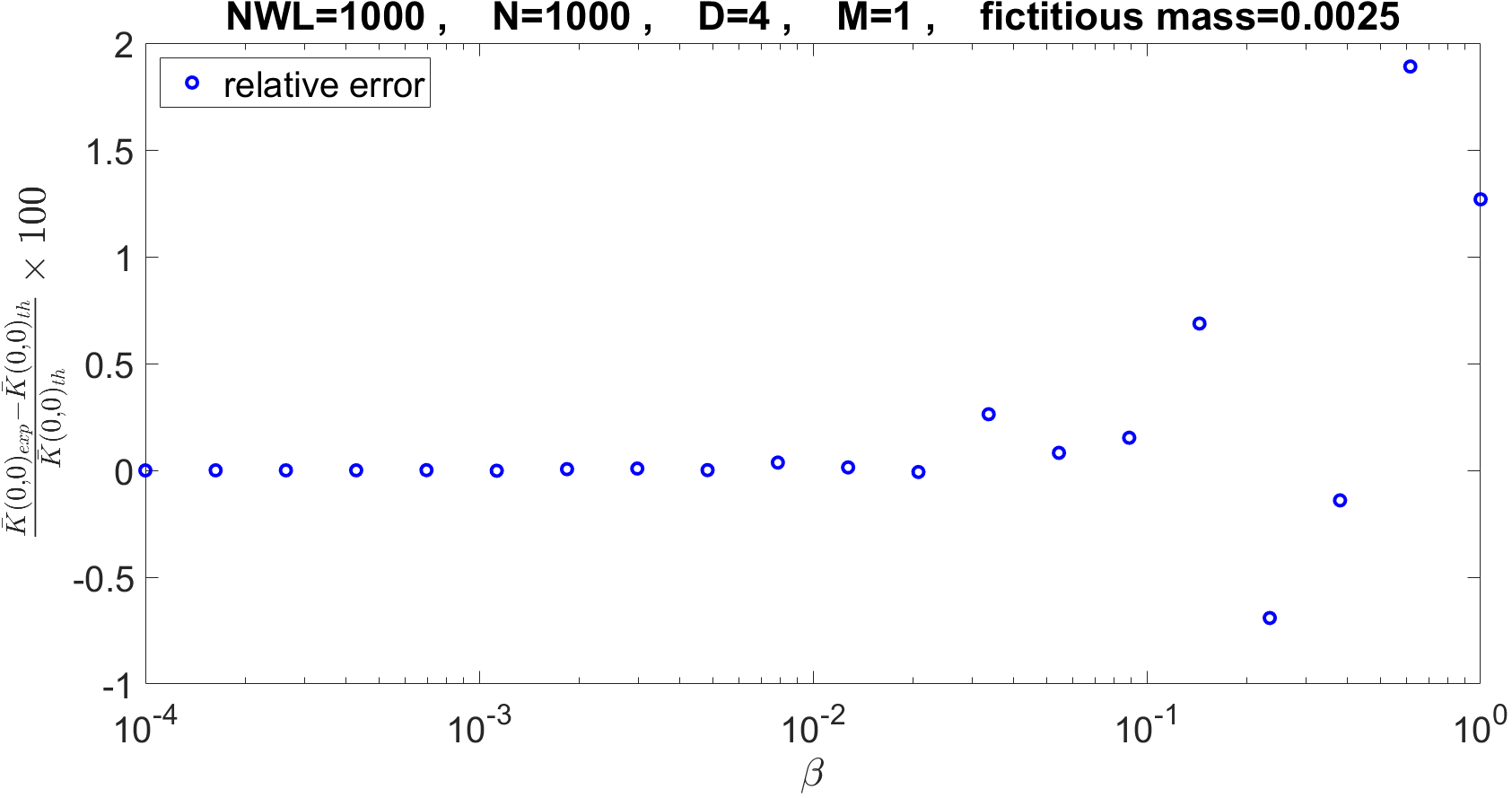}
        }\\
        \subfigure[]{%
            \label{fig:M_0p1_prop}
            \includegraphics[width=0.5\textwidth]{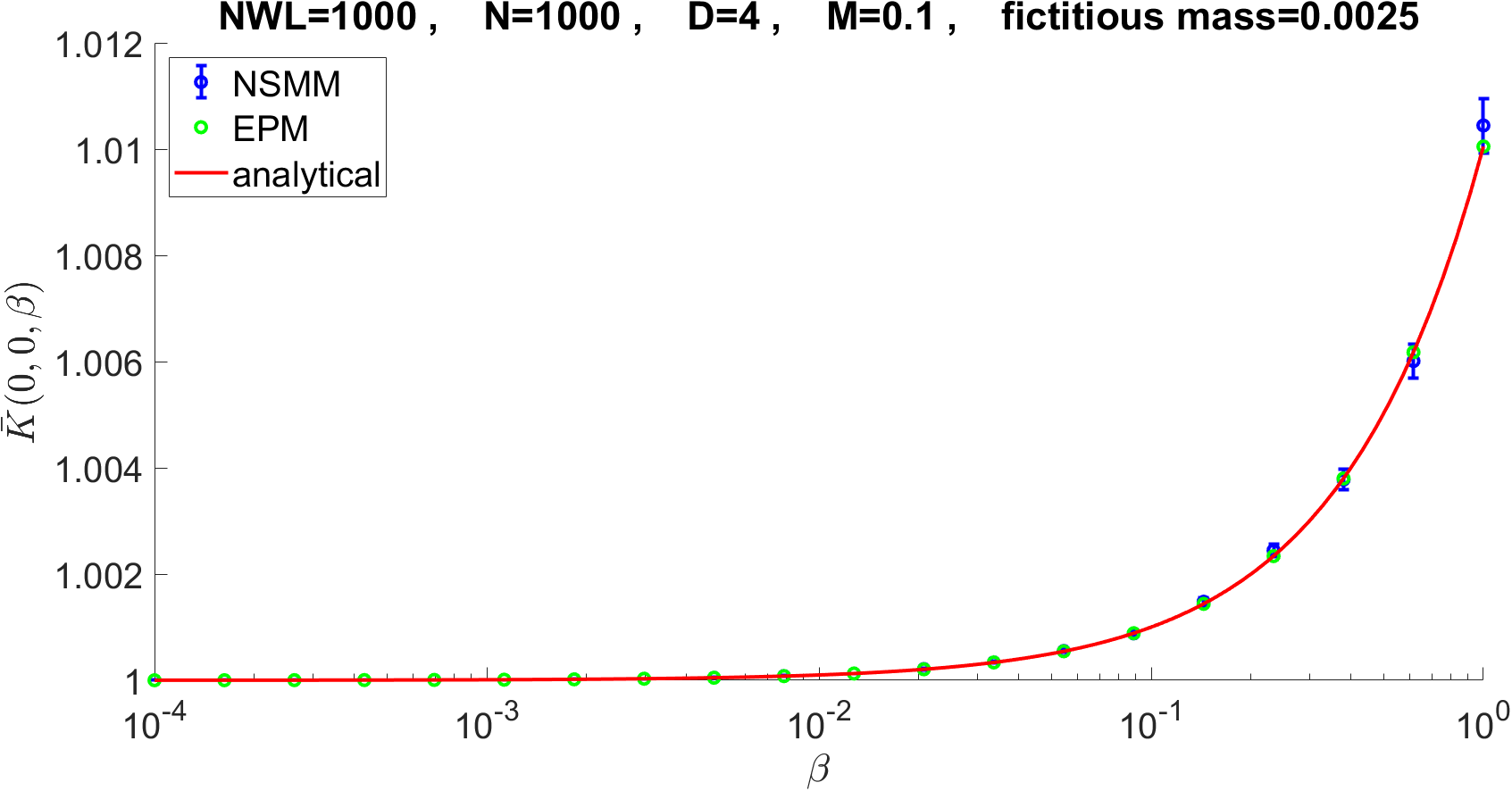}
        }%
        \subfigure[]{%
           \label{fig:M_0p1_relerr}
           \includegraphics[width=0.5\textwidth]{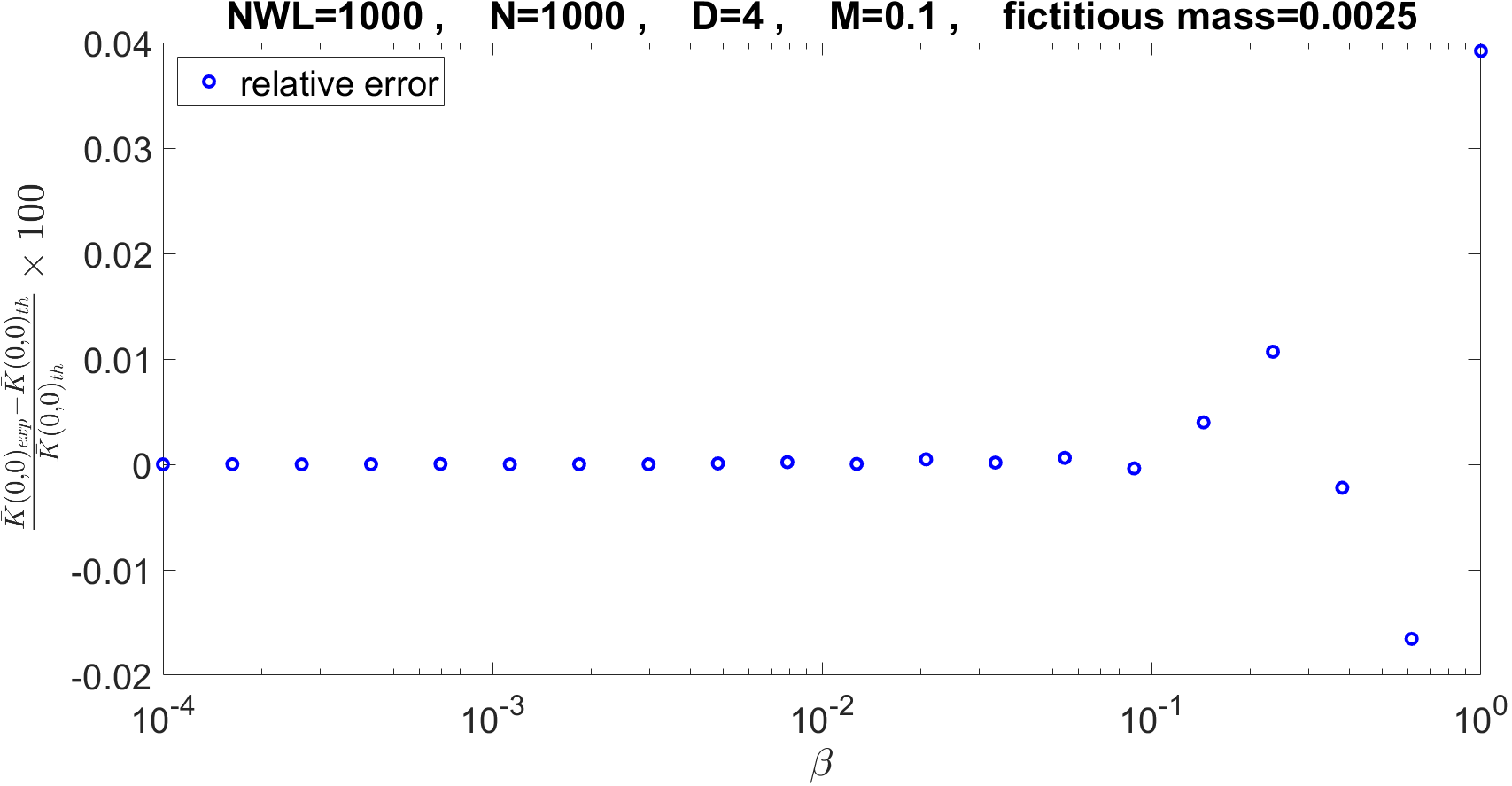}
        }
    
    \caption{%
        Calculations of the heat kernel for the free scalar particle on the sphere with $M=1$ (top) and $M=0.1$ (bottom). The relative errors between the blue and the green data have been shown in the figures on the right. All other parameters are kept fixed.
     \label{fig:prop_relerr_M}}%
\end{figure}
Next, in Figures~\ref{fig:prop_relerr_M}, we show the comparison between the two calculations performed for different values of the curvature parameter $M$ (the inverse radius of the sphere), namely $M=1$ and $M=0.1$, reporting also the relative errors between the two methods. What we observe is that, locally, the decrease of curvature improves the agreement between NSMM and EPM---the precision has gained roughly two orders of magnitude (the full-scale changes from $2\%$ to $0.04\%$) passing from $M= 1$ to $M= 0.1$. This is due to the fact that, magnifying the radius of the sphere, the space in the vicinity of $x = x' = 0$ gets closer to be flat, reducing the stronger effects which affect the precision of the NSMM.

In general, the different degree of precision between the two methods considered, can be ascribed to the fact that---as already observed in perturbative computations~\cite{Bastianelli:2017wsy}---the flat effective model (EPM) takes into account of the curvature effects in a much more efficient way than the non-linear sigma model, as in the former the curvature effects are all encoded in the effective potential. In fact, in the non-linear sigma-model, derivative interactions do appear, which are less efficiently represented in the heat-kernel expansion. This can be understood, by rewriting the action~\eqref{eq11} in terms of a rescaled time $s=\tau/\beta$ which yields
\begin{align}
    S[x] =\int_0^1ds \Biggl( \frac{1}{2\beta} g_{\mu\nu} \dot x^\mu \dot x^\nu +\beta \Big(V_{CT}(x)+V(x)\Big)\Biggr)\,,
\end{align}
in which, the potentials appear with an order-$\beta^2$ higher than the derivative terms. Thus, in a perturbative expansion about a fixed point, one needs higher order terms from the Taylor expansion of $g_{\mu\nu}$ to match the corresponding orders in the potential. On the other hand, the flat effective model---unlike the non-linear sigma model---has, so far, been shown to work only in maximally symmetric backgrounds.   

\begin{figure}[ht!]
   \centering
            
            \includegraphics[width=0.7\textwidth]{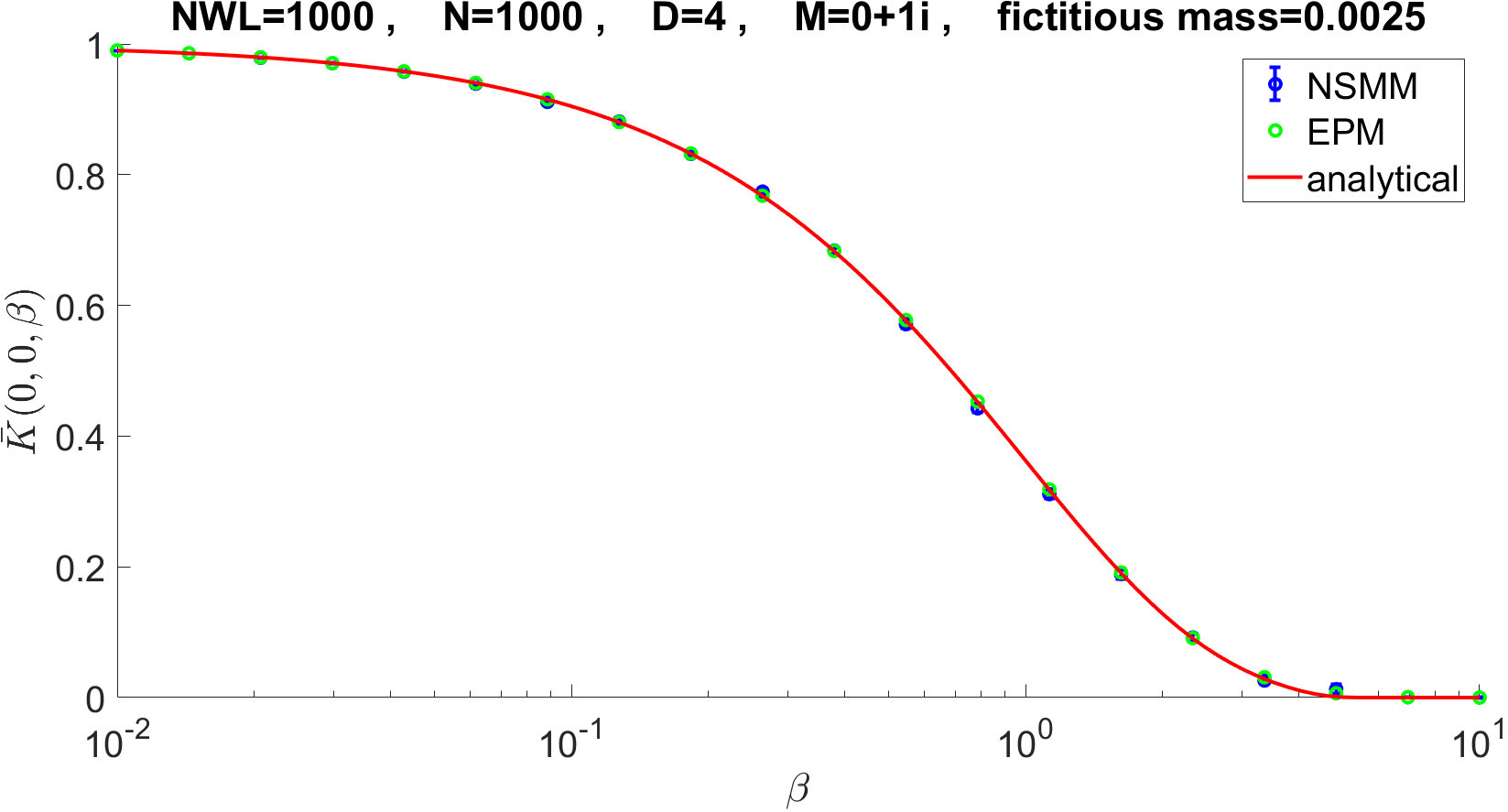}
    \caption{%
        Calculations of the heat kernel for the free scalar particle on a {\color{black}hyperboloid} with $D=4$ and $\left\lvert M\right\rvert=1$.}
\label{fig:M_1i_prop}
\end{figure}

Finally, we present the calculation of the free particle heat kernel constrained on a {\color{black}maximally symmetric space with negative curvature} (a 4-hyperboloid in our case). Following \cite{Bastianelli:2017wsy}, the sectional curvature is negative $M^2<0$ and, defining $\left\lvert M \right\rvert=\sqrt{-M^2}$, the metric is written as 
\begin{equation}
    g_{\mu\nu}(x)=\delta_{\mu\nu}+\tilde f(x)P_{\mu\nu}(x)
\end{equation}
with
\begin{equation}
    \tilde f(x)=\frac{-1-2\left( \left\lvert M\right\rvert x\right)^2 + \cosh{\left(2\left\lvert  M\right\rvert x\right)}}{2\left( \left\lvert M\right\rvert x \right)^2}.
\end{equation}
All other quantities appearing in \eqref{eq34} are defined in the same way as before, whereas in the potential \eqref{eq20} we have the replacement $M\to iM$. Figure \ref{fig:M_1i_prop} shows a satisfying agreement between the curved method and the flat one for $(N_{WL},N)=(1000,1000)$ for $\beta$-values ranging from 0.01 to 10.

\section{An extension to open worldlines}
\label{sec:openWL}
So far we have seen an efficient way to numerically compute particle path integrals in curved space, with a direct application to the case of the heat kernel of a free scalar particle on a maximally symmetric space with Dirichlet boundary conditions at the origin, $x(0)=x(\beta)=0$. In this section we present a possible way to easily extend the above computation to the case of non-zero boundary conditions. First of all, let us denote the endpoints with $x(0)=y$ and $x(\beta)=z$ for the particle $x(\tau)$ propagating from time $\tau=0$ to time $\tau=\beta$. The path integral we are interested in, is
\begin{equation}
\label{eq35}
I(y,z;\beta)=\smashoperator{\int_{x(0)=y}^{x(\beta)=z}}\mathcal Dx\;e^{-S[x]},
\end{equation}
with
\begin{equation}
\mathcal Dx=\smashoperator{\prod_{\tau=0,\ldots,\beta}}\sqrt{g\left(x(\tau)\right)}dx(\tau),\quad S[x]=\int\displaylimits_0^{\beta}d\tau\left( \frac{1}{2}g_{\mu\nu}(x)\dot x^{\mu} \dot x^{\nu}+\tilde V(x) \right),
\end{equation}
where the potential $\tilde V(x)$ includes the counter-term \eqref{eq21} and a possible external potential. Now, we perform a splitting of the particle path into the classical one $x_{cl}(\tau)$ (\textit{i.e.} a straight line, being the solution of the classical equation of motion for the free particle) and quantum fluctuations $q(\tau)$,
\begin{equation}
x^{\mu}(\tau)=x_{cl}^{\mu}(\tau)+q^{\mu}(\tau),\quad x_{cl}^{\mu}(\tau)=y^{\mu}+\frac{\tau}{\beta}\left(z^{\mu}-y^{\mu}\right).   
\end{equation}
In particular, the fluctuations satisfy $q(0)=q(\beta)=0$. Hence, the average of the path integral \eqref{eq35} takes the form
\begin{equation}
\label{eq36}
\left\langle I(y,z;\beta)\right\rangle = e^{-\frac{\left( y-z \right)^2}{2\beta}}\left\langle\quad\smashoperator{\int_{q(0)=0}^{q(\beta)=0}}\mathcal Dq\;e^{-S_{KIN}[x_{cl},q]-S_{POT}[x_{cl},q]}\right\rangle,
\end{equation}
with
\begin{gather}
\mathcal Dq=\smashoperator{\prod_{\tau=0,\ldots,\beta}}\sqrt{g\left(x_{cl}(\tau)+q(\tau)\right)}dq(\tau),\\
\label{eq38}S_{KIN}[x]=\int\displaylimits_0^{\beta}d\tau\left[ \frac{1}{2}\left(g_{\mu\nu}(x_{cl}+q)-\delta_{\mu\nu}\right)\left( \frac{1}{\beta}(z^{\mu}-y^{\mu})+\dot q^{\mu} \right)\left( \frac{1}{\beta}(z^{\mu}-y^{\mu})+\dot q^{\mu} \right)\right],\\
S_{POT}[x]=\int\displaylimits_0^{\beta}d\tau\left[\tilde V(x_{cl}+q)\right].
\label{eq37}
\end{gather}
As equations \eqref{eq36}-\eqref{eq37} show, it is possible to calculate the averaged path integral $\left\langle I(y,z;\beta) \right\rangle$ with non-vanishing endpoints in terms of another one with null Dirichlet boundary conditions (like those computed in Section \ref{sec:CS_fsHToS}), provided a few replacements/adjustments, namely:
\begin{itemize}
    \item a factor which depends upon the squared difference of the endpoints appears in front of the average~\eqref{eq36}---which corresponds to the free flat action;
    \item the sampled points of the worldlines around zero (which here are denoted by $q$) must be displaced by the corresponding points of the classical path $x_{cl}$ when we evaluate the potentials \eqref{eq38} and \eqref{eq37}.
\end{itemize}

\begin{figure}[ht!]
   \centering
            
            \includegraphics[width=0.7\textwidth]{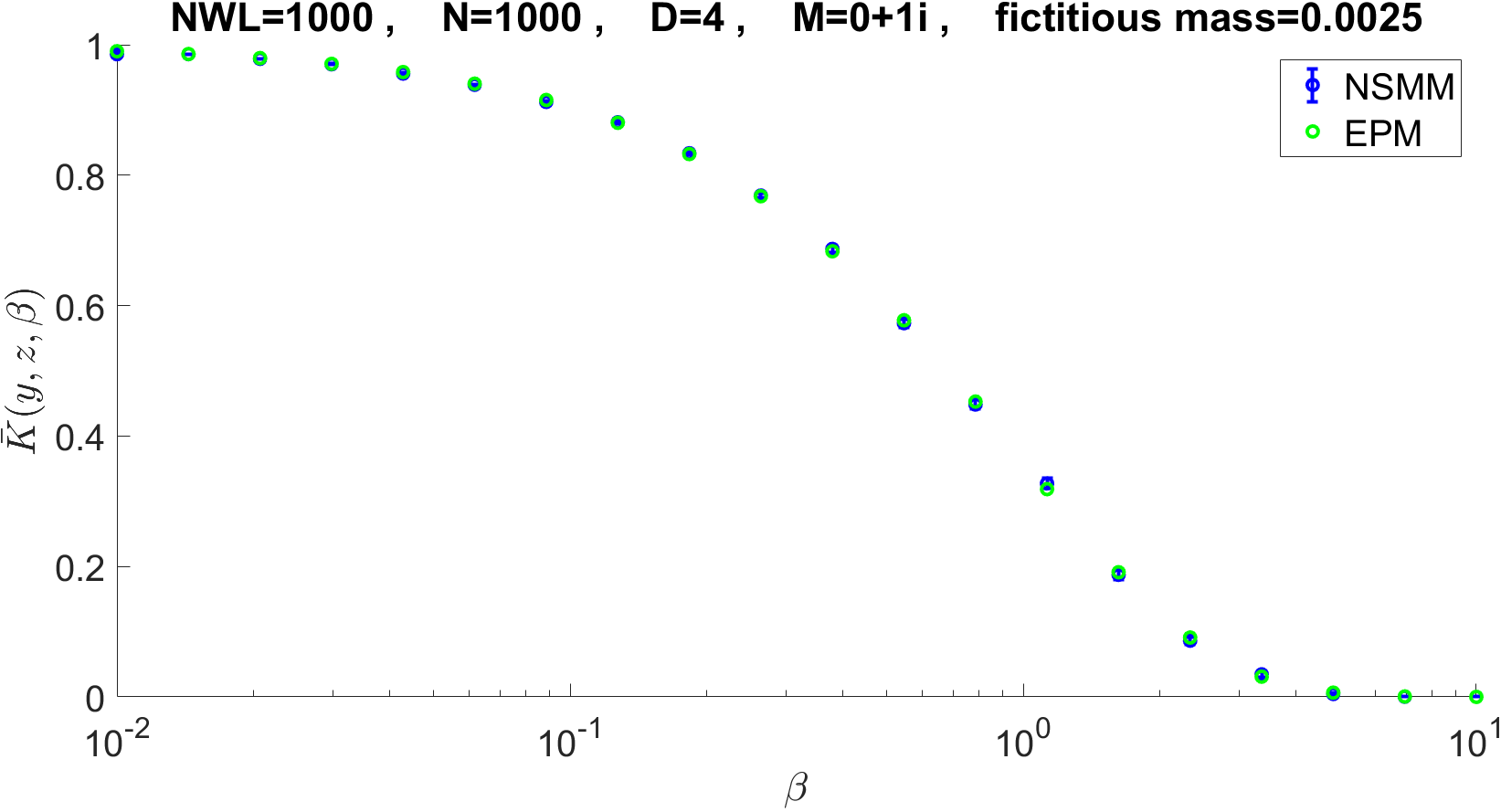}
    \caption{%
        Calculations of the heat kernel for the free scalar particle propagating from $y=(0,0,0,0)$ to $z=(5,5,5,5)$ on a {\color{black}hyperboloid} with $D=4$ and $\left\lvert M\right\rvert=1$. The blue points show a calculation performed with the non-linear sigma model method, whereas the green points are referred to a benchmark calculation obtained with the flat space setup.}
\label{fig:openWL_prop}
\end{figure}

In Figure \ref{fig:openWL_prop} we report the calculation of the heat kernel of a free scalar particle propagating on a maximally symmetric space with $D=4$, from point $y=(0,0,0,0)$ to point $z=(5,5,5,5)$, done using both numerical methods introduced above, and showing very good agreement. 


\section{Conclusions and Outlook}
\label{sec:C}
In the past couple of decades the numerical Monte Carlo approach to the Worldline formalism in flat space has seen a variety of applications. In the meantime, worldline techniques have been extended to QFT in curved spaces, but restricted to perturbative, analytic computations. The present manuscript intends to be a first attempt to link these two avenues, in order to be able to extend the Worldline approach to QFT in curved space beyond analytic perturbation theory. Applications can be numerous. Firstly, the study of quantum field theory effective actions can be extended beyond the regime where the inclusion of gravity is treated perturbatively, as for instance in the derivation of gravitational corrections to the Euler-Heisenberg lagrangians~\cite{Bastianelli:2008cu}. On the other hand, the effect of curvatures in strongly-coupled fermionic models, such as the Gross-Neveu model, which describe the low-energy limit of several physical systems, can be studied with great efficiency with the present setup. In fact, in flat space, the Worldline Monte Carlo approach to large $N$ fermionic models was already successfully considered~\cite{Dunne:2009zz,Mazur:2014hta}, whereas, at the perturbative level, the heat kernel expansion of the Gross-Neveu model in 3d curved space with constant curvature was studied, for example in Ref.~\cite{Miele:1996rp}. The results that we have presented here clearly show the possibility of performing worldline path integrals associated to scalar point particles moving in curved space by means of Monte Carlo numerical procedures. 

In the present manuscript, we have considered maximally symmetric geometries. By using a recently studied {\it effective} flat space model---which encodes the curvature effects inside a suitable potential---as a benchmark, we have described a possible extension of the Worldline Monte Carlo to curved spaces with maximal symmetry. We have found it convenient to add a mass term in the kinetic action that serves to construct the worldline ensemble. Let us stress that, although the computation presented above focuses on maximally symmetric spaces, in the numerical NSMM approach there is {\it a priori} no particular constraint on the curved space background, nor is there any limitation on the chosen set of coordinates. However, as in other numerical methods, the main drawback of the present approach is the necessity of having a specific background, dependent on a finite set of real parameters. 

The construction described in the present manuscript focuses on the computation of the diagonal part of a particle heat kernel which is linked to the computation of  one-loop effective actions. However, a possible strategy to simulate {\it open worldlines} has been presented and might also be used to study dressed particle propagators, for instance. Finally, the inclusion of spinorial degrees of freedom would certainly be a welcome generalization. This problem can be tackled either with the inclusion of suitable matrix-valued potentials (spin factors) or in terms of spinning particle models, which involve Grassmann odd coordinates, along with the geometric, Grassmann even, coordinates. Numerical approaches in the presence of Grassmann odd variables were already considered in \cite{Creutz:1998bg}, and it would be helpful to apply such construction to the present worldline Monte Carlo method.

\section*{Acknowledgments}
We would like to thank Christian Schubert and James P. Edwards for fruitful discussions and directions about the use of their numerical algorithms, that we have adopted in the present work. We thank also Antonino Flachi and So Matsuura for the very useful suggestions, especially concerning the insertion of a mass contribution in the WL sampling procedure.


\appendix

\section{Modified \textit{YLOOPS} algorithm}
\label{app:A}
To diagonalize the flat kinetic term \eqref{eq2}, we first define the discretized derivative as the \textit{backward difference}
\begin{align}
\label{eq14}
\dot x^{\mu}(\tau)\simeq\frac{N}{\beta}\left( x_k-x_{k-1} \right).
\end{align}
The index $k$ parameterizes the discretized $x$-points in the same way as the continuous parameter $\tau$ parameterizes the continuous $x$-points. Now, the full kinetic term will be proportional to the summation
\begin{align}
Y^{(0)}=\sum_{k=2}^N\left(x_k-x_{k-1}\right)^2,
\end{align}
which we generalize to
\begin{align}
\label{eq22}
Y^{(\alpha)}=\sum_{k=2}^N\left[\left(x_k-x_{k-1}\right)^2+\alpha x_k^2\right],\quad\alpha>0.
\end{align}
Following the same steps of \cite{Edwards:2019fjh},
\begin{align}
\label{eq32}
Y^{(\alpha)}=&\sum_{k=2}^N\left[\left(x_k-x_{k-1}\right)^2+\alpha x_k^2\right]=\sum_{k=3}^{N-1}\left[\left(x_k-x_{k-1}\right)^2+\alpha x_k^2\right]+x_2^2+x_{N-1}^2+\alpha x_2^2\nonumber\\
=&\sum_{k=1}^{N-2}C^{(\alpha)}_k\left( x_{N-k}-\frac{1}{C^{(\alpha)}_k}x_{N-k-1} \right)^2\nonumber\\
=&\sum_{k=2}^{N-1}C^{(\alpha)}_{N-k}\bar x_k^2,
\end{align}
where
\begin{align}
\label{eq33}
\bar x_k=x_k-\frac{1}{C^{(\alpha)}_{N-k}}x_{k-1}
\end{align}
and
\begin{align}
\label{eq31}
& C^{(\alpha)}_1=2+\alpha\nonumber\\
& C^{(\alpha)}_k=C^{(\alpha)}_1-\frac{1}{C^{(\alpha)}_{k-1}}, \quad k=2,\ldots,N-2.
\end{align}
The algorithm finally reads as follows.
\begin{enumerate}
\item Generate $N-2$ vectors $\omega_i,\; i=1,\ldots,N-2$ distributed according to $\frac{1}{\left( 2\pi \right)^{D/2}}e^{-\omega_i{}^2}$.
\item Compute
\begin{align}
\bar x_i=\sqrt{\frac{2}{N}\frac{1}{C^{(\alpha)}_{N-i}}}\omega_{i-1},\quad i=2,\ldots,N-1.
\end{align}
\item Construct the loop according to
\begin{align}
x_1=&x_{N}=0\nonumber\\
x_2=&\bar x_2\nonumber\\
x_i=&\bar x_i+\frac{1}{C^{(\alpha)}_{N-i}}x_{i-1},\quad i=3,\ldots,N-1.
\end{align}
\end{enumerate}
It is easy to see that coefficients $C^{(\alpha)}_k$ reproduce those of \cite{Edwards:2019fjh} for vanishing $\alpha$, i.e. $\frac{k+1}{k}$.

\section{WLMC convergence with tiny mass term}
\label{app:B}
The effect of adding a tiny mass term to the sampling algorithm of the worldlines (cfr. Eq. \eqref{eq22}) has an important effect on WLMC computations.
\begin{figure}[ht!]
    \begin{center}
        \subfigure[]{%
            \label{fig:ym1}
            \includegraphics[width=0.5\textwidth]{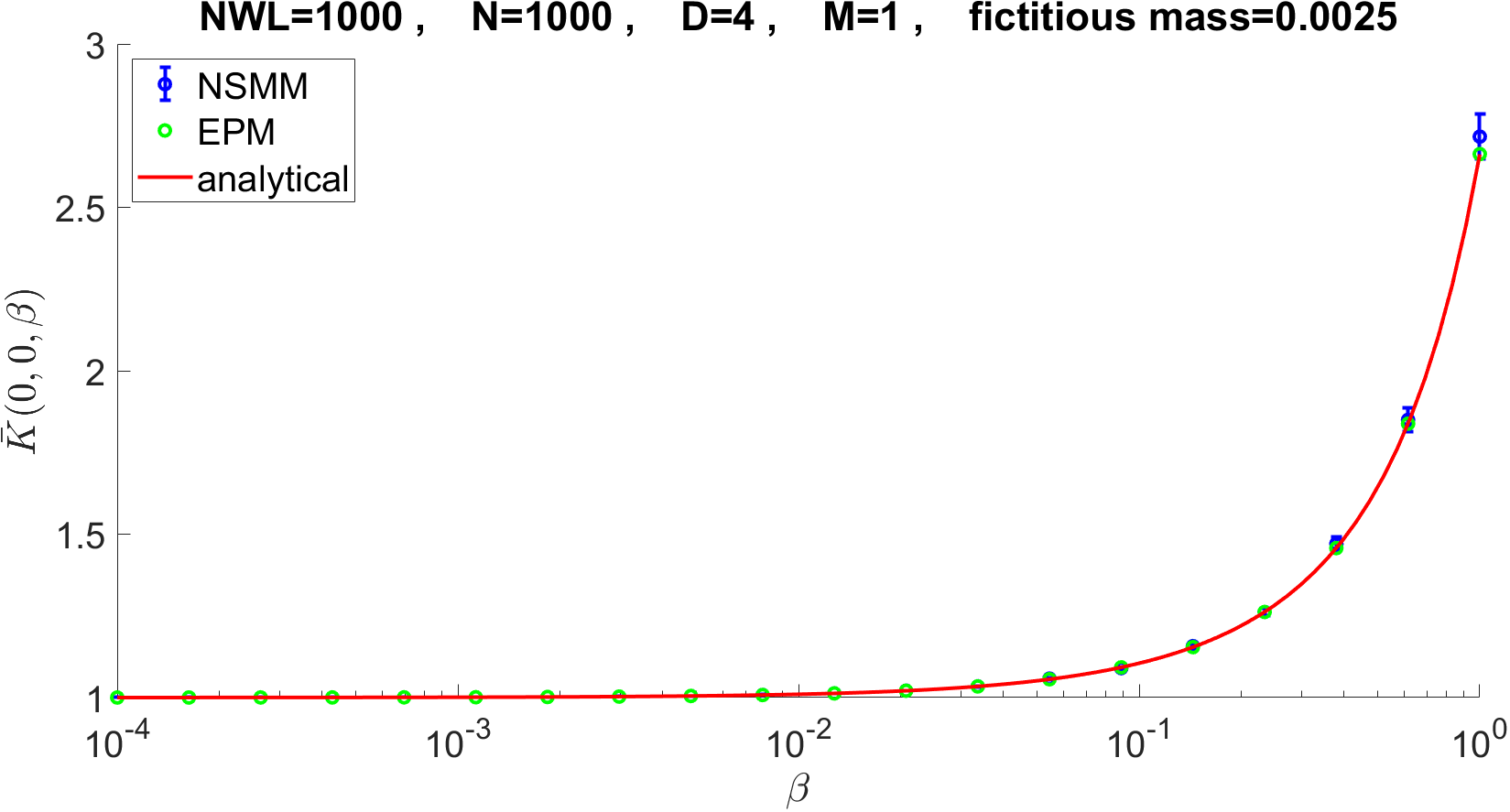}
        }%
        \subfigure[]{%
            \label{fig:nm1}
            \includegraphics[width=0.5\textwidth]{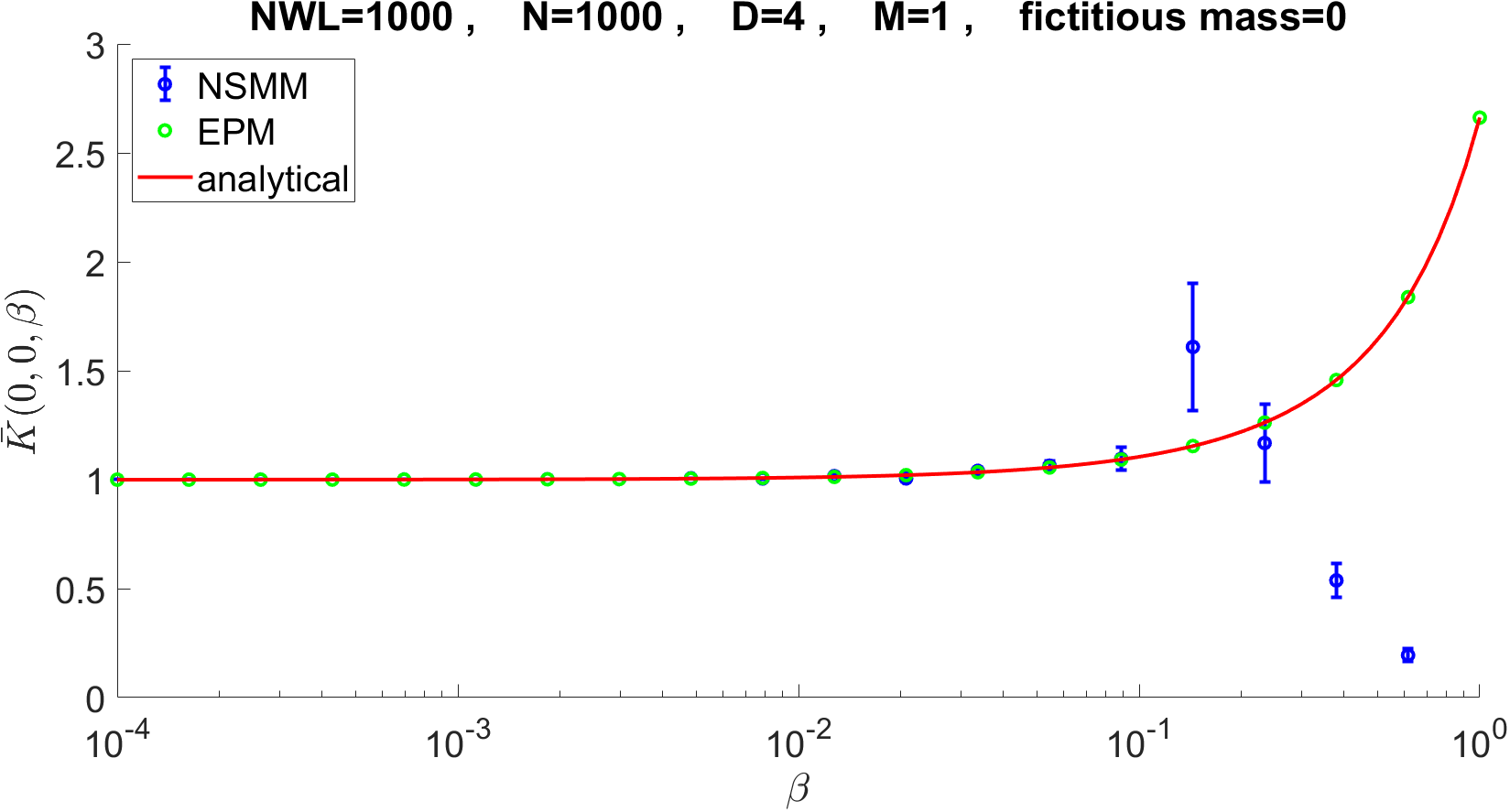}
        }%
    \end{center}
    \caption{%
        Two calculations of the heat kernel of a free scalar particle on a 4-sphere with $(N_{WL},N)=(1000,1000)$ and $M=1$. For the left figure, a tiny fictitious mass has been introduced ($\alpha=0.0025$) in the WL sampling. The right figure is obtained with ($\alpha=0$), i.e. no fictitious mass term.
     }%
   \label{fig:mass_conv}
\end{figure}
In Figure \ref{fig:mass_conv}, we report the case of two calculations of the heat kernel introduced in Section \ref{sec:CS_fsHToS}. Both calculations have been performed with the same parameters, namely $N_{WL}=1000$, $N=1000$, $D=4$ and $M=1$: however, for the one on the left a tiny fictitious mass has been introduced in the sampling, whilst for the other it has not. It can be easily noticed that \ref{fig:ym1} follows with good accuracy the expected behaviour, while \ref{fig:nm1} not only has a lot more broadened data, but they also qualitatively deviate from the green reference points. In this context, the addition of a mass term has the role of a \textit{regulator} for the sampling of the worldlines. We stress that this fictitious mass has not been taken into account during the \textit{evaluation of the potential}; rather, its only role is to modify the diagonalization coefficients
\eqref{eq31} appearing in \eqref{eq32} and \eqref{eq33}.

Finally, we investigated a possible qualitative dependence of the fictitious mass which minimizes the error between the non-linear sigma model method and the effective potential method, with respect to the curvature of the maximally symmetric space. We considered the hyperboloid case with $\lvert M \rvert$ ranging from 1 to 0.01. However, nothing would forbid to consider the positive curvature case, and the results are expected to hold unchanged. 

\begin{figure}[ht!]
    \begin{center}
        \subfigure[]{%
            \label{fig:m_vs_curv_1}
            \includegraphics[width=0.5\textwidth]{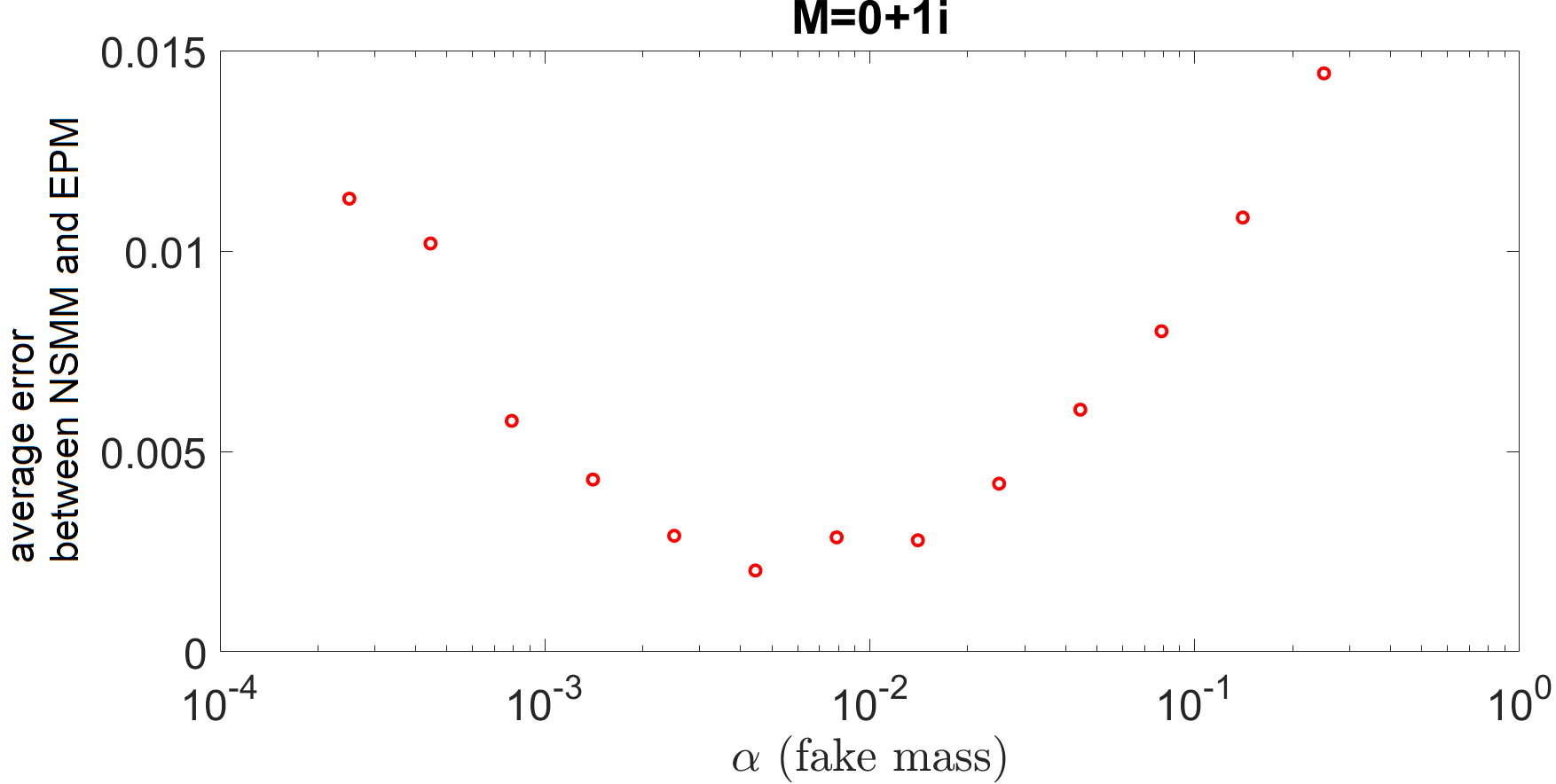}
        }%
        \subfigure[]{%
            \label{fig:m_vs_curv_0p01}
            \includegraphics[width=0.5\textwidth]{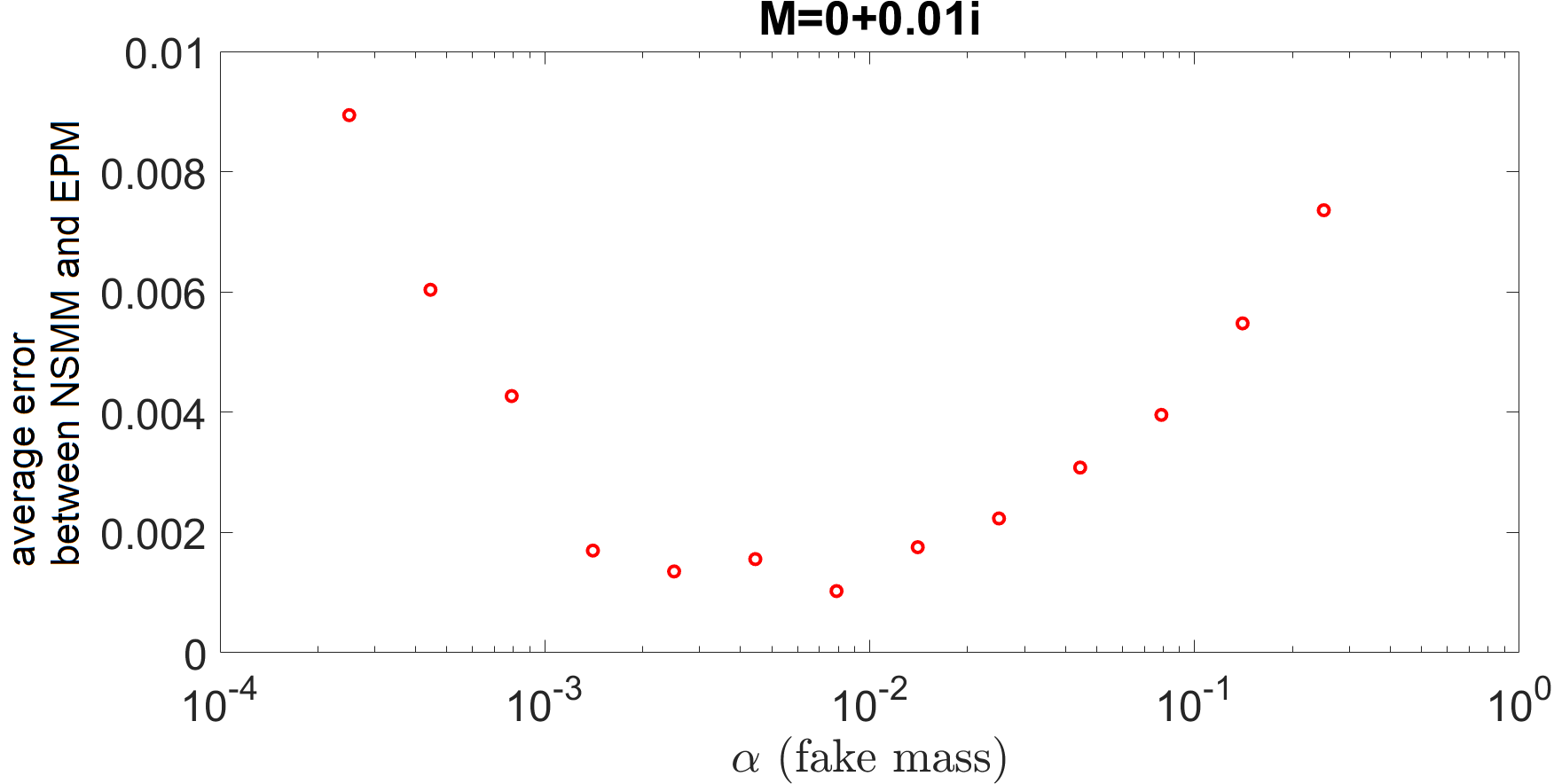}
        }%
    \end{center}
    \caption{%
        Comparison between heat kernel calculations of a free scalar particle on the hyperboloid for $\lvert M \rvert=1,\;0.01$. Each point denotes a single simulation: on the horizontal axis we have the value of the fictitious mass parameter $\alpha$ adopted, whereas vertically we have reported an arithmetic average of distance between the data of the non-linear sigma model method and the ones of the effective potential method (assumed as a benchmark). The $\alpha$-value which minimizes the error does not qualitatively depend on the curvature of the space.
     }%
   \label{fig:mass_vs_curv}
\end{figure}

\begin{figure}[ht!]
   \centering
            
            \includegraphics[width=0.7\textwidth]{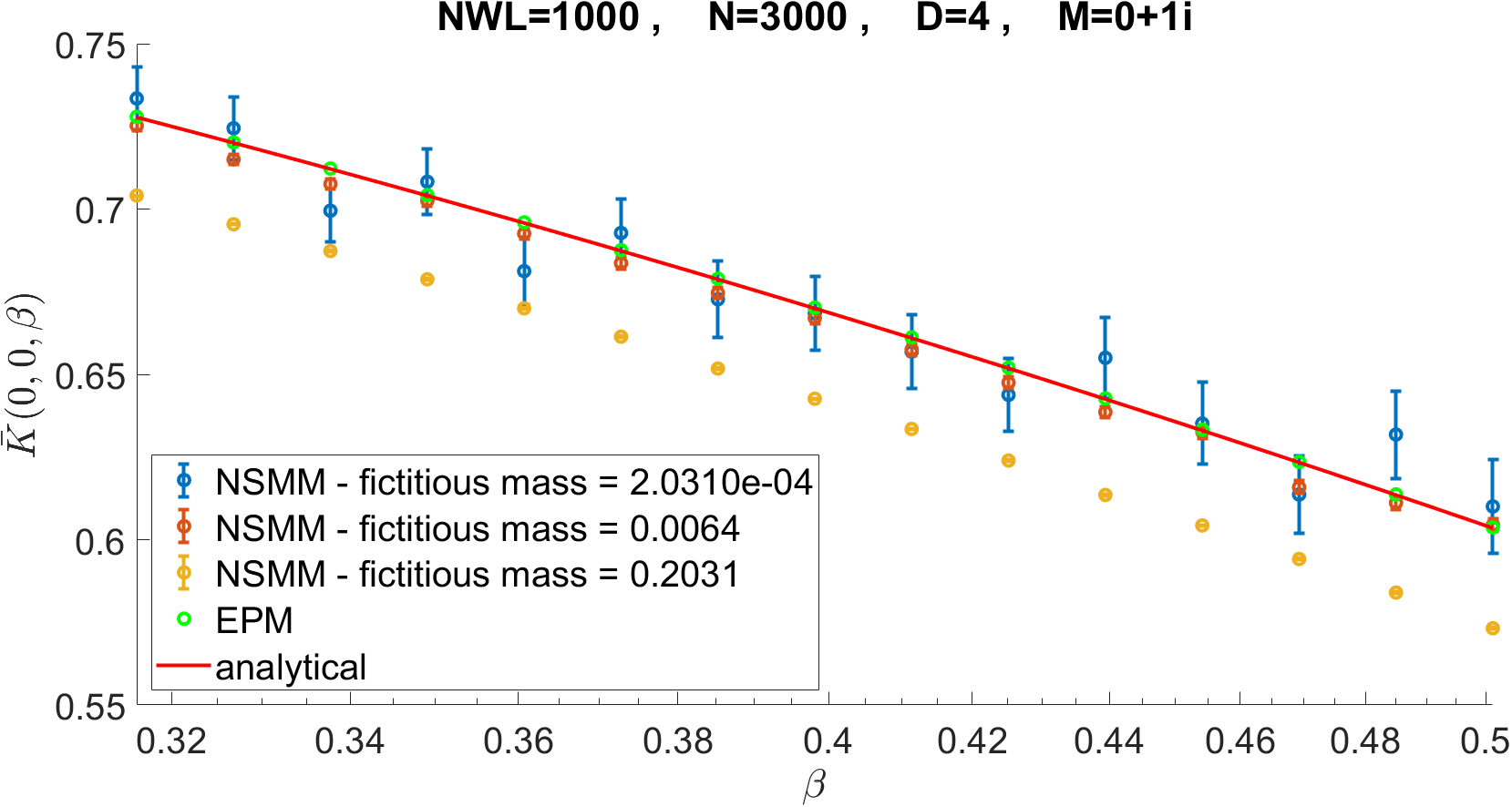}
    \caption{%
        Calculations of the heat kernel for the free scalar particle on a hyperboloid with $D=4$ and $\left\lvert M\right\rvert=1$, for a window of $\beta$-values. The plot shows the comparison of the calculation when different values of $\alpha$ are implemented, in particular $\alpha_{min}=0.0064$, $\alpha_{-}=0.0002$ and $\alpha_{+}=0.2031$.}
\label{fig:im3}
\end{figure}

In Figure \ref{fig:mass_vs_curv} we report the cases of $\lvert M \rvert=1$ and $\lvert M \rvert=0.01$ (intermediate curvatures exhibit the same behaviour). Each point of the plot represents the average error between a simulation with the non-linear sigma model method and the associated effective potential method for different values of the $\alpha$ parameter. It can be seen that the optimization value $\alpha_{min}$ does not change over two orders of magnitude of the curvature of the space. Hence we conclude that the fictitious mass parameter $\alpha$ is substantially curvature-independent. {\color{black}It is possible to see the minimized average error directly at work: let us consider the left panel of Figure \ref{fig:mass_vs_curv} ($M=i$) and extrapolate the minimum through a quadratic fitting (still on the log-scale), obtaining $\alpha_{min}=10^{\bar{\alpha}}$. Then we consider two logarithmically equally spaced values, \textit{i.e.} $\alpha_{-}=10^{\bar{\alpha}-\Delta}$ and $\alpha_{+}=10^{\bar{\alpha}+\Delta}$, with $\Delta>0$. In Figure \ref{fig:im3} we adopted $\Delta=1.5$ (arbitrary) and $\bar{\alpha}=2.1923$ (logarithmic minimum fitted for the left panel of Figure \ref{fig:mass_vs_curv}) and computed a portion of the propagator of the free particle on the hyperboloid: as expected the red data (corresponding to $\alpha_{min}$) are those which better reproduce the expected behaviour of the propagator.}

\end{document}